[1994/12/01]
\documentclass[final]{article}
\AtEndDocument{\bigskip{\footnotesize%
  \noindent $^1$ \textsc{Sodankyl\"a Geophysical Observatory, University of Oulu, FI-90014 Oulu, Finland}.  \textit{E-mail address}: \texttt{lassi.roininen@oulu.fi} \par
  \addvspace{\medskipamount}
\noindent   $^2$ \textsc{Department of Mathematics and Statistics, University of Helsinki, FI-00014 Helsinki, Finland}. \par
\noindent \textit{E-mail address}: \texttt{petteri.piiroinen@helsinki.fi}\par
  \addvspace{\medskipamount}
  \noindent   $^3$ \textsc{Deka Investment GmbH, 60325 Frankfurt am Main, Germany}. \par 
  \noindent T.~Schoden \textit{E-mail address}: \texttt{t.schoden@yahoo.de}\par \noindent M.~Simon \textit{E-mail address}: \texttt{info@simon-martin.net}
  
}}

\title{Asset Price Bubbles: An Option-based Indicator}
\author{Petteri Piiroinen$^1$, Lassi Roininen$^2$, Tobias Schoden$^3$\\ and Martin Simon$^3$ \\ \\ $^1$University of Helsinki, Finland \\ $^2$University of Oulu, Finland \\ $^3$Deka Investment GmbH, Germany}
\date{Version 1.0, \today}

\usepackage{amsmath,amsthm}
\usepackage{amssymb}
\usepackage{graphicx}
\usepackage{subcaption}
\usepackage{amssymb}
\usepackage{amsfonts}
\usepackage{tikz}
\usepackage{verbatim}
\usepackage{algorithm,algorithmic}
  \usepackage{placeins}

\chardef\bslash=`\\ 

\newtheorem{thm}{Theorem}[section]

\theoremstyle{defn}
\newtheorem{defn}{Definition}[section]

\theoremstyle{rem}
\newtheorem{rem}{Remark}[section]
\newtheorem{ex}{Example}[section]

\newcommand{\Eb}{\boldsymbol{E}}

\newcommand{\e}{\boldsymbol{e}}

\newcommand{\dd}{\,\mathrm{d}}
\newcommand{\EW}{\mathbb{E}}

\newcommand{\eval}[2][\right]{\relax
  \ifx#1\right\relax \left.\fi#2#1\rvert}

\newcommand\blfootnote[1]{%
  \begingroup
  \renewcommand\thefootnote{}\footnote{#1}%
  \addtocounter{footnote}{-1}%
  \endgroup
}

\begin{document}
\maketitle
\markboth{Asset Price Bubbles: An Option-based Indicator}
{Asset Price Bubbles: An Option-based Indicator}
\renewcommand{\sectionmark}[1]{}

\blfootnote{The opinions expressed in this article are those of the authors and do not necessarily reflect views of Deka Investment GmbH.}
\blfootnote{This work has been funded by Academy of Finland (decision
numbers 284715, 307741, 312119 and 313709).}
\blfootnote{MS is indebted to Dominik Garmatter for valuable discussions of this material.}

\begin{abstract}
We construct a statistical indicator  for the detection of short-term asset price bubbles based on the information content of bid and ask market quotes for plain vanilla put and call options. 
Our construction makes use of the martingale theory of asset price bubbles and the fact that such scenarios where the price for an asset exceeds its fundamental value can in principle be detected by analysis of the asymptotic behavior of the implied volatility surface.
For extrapolating this implied volatility, we choose the SABR model, mainly because of its decent fit to real option market quotes for a broad range of maturities and its ease of calibration. As main theoretical result, we show that under lognormal SABR dynamics, we can compute a simple yet powerful closed-form martingale defect indicator by solving an ill-posed inverse calibration problem. 
In order to cope with the ill-posedness and to quantify the uncertainty which is inherent to such an indicator, we adopt a Bayesian statistical parameter estimation perspective.
We probe the resulting posterior densities with a combination of optimization and adaptive Markov chain Monte Carlo methods, thus providing a full-blown uncertainty estimation of all the underlying parameters and  the martingale defect indicator.
Finally, we provide real-market tests of the proposed option-based indicator with focus on tech stocks due to increasing concerns about a tech bubble 2.0.
\end{abstract}

\section{Introduction}
The martingale theory of asset price bubbles has attracted considerable interest both from the theoretical and practical point of view over the last years and there is now a vast body of literature on the topic. An asset price bubble appears when the \emph{market value} of an asset exceeds its \emph{fundamental value}. The authors of the works \cite{CoxHobson,Tysk0,Tysk1,Tysk2, Heston, Jarrow1,Jarrow2,Kardaras,Loewenstein1, Loewenstein2, Protter1} propose to use a discounted underlying asset price process which is a strict local martingale, i.e., a local martingale but not a martingale, in order to model asset price bubbles. We refer to Protter ~\cite{Protter} for an excellent review of the martingale theory of asset price bubbles. For a study of the impact an asset price bubble has on standard risk management methodologies we refer the reader to Jarrow~\cite{Jarrow}.

Practical asset price bubble detection in strict local martingale frameworks has been studied previously by Jarrow, Protter and their collaborators in the series of papers \cite{Jarrow3,Jarrow4,Jarrow5,Jarrow6}; their numerical indicators rely on asset price time series in a one-factor time-homogeneous local volatility model for the underlying. More recently, the fact that stock options are liquidly traded for a wide range of strikes has been identified as a promising path for defining novel indicators for asset price bubbles, see Jarrow~\cite{Jarrow6}. Such an option-based indicator in terms of option-implied volatility has been introduced by Jacquier and Keller-Ressel in \cite{Jacquier}, showing that bubbles can in principle be detected by analysis of the asymptotic behavior of the implied volatility surface. 

In this work, we introduce a \emph{statistical} indicator for the detection of asset price bubbles based on the information content of bid and ask market quotes for the prices of plain vanilla put and call options. More precisely, we embed our \emph{martingale defect indicator} into a statistical framework thus enabling  both, providing an indication whether the discounted underlying stock price process may be modeled as a strict local martingale and quantifying the uncertainty inherent to such an indicator. In contrast to the aforementioned previous works, this new approach quantifies the severity of the supposed asset price bubble which is particularly appealing for risk management purposes both at buy and sell side market participants. While the statistical framework introduced in this work is model-independent, we choose the SABR stochastic volatility model and we argue that for the purpose of detecting short-term bubbles this is both sufficiently accurate and sufficiently simple to enable straight forward integration into existing risk management infrastructure. Finally, we would like to emphasize that to the best of the authors' knowledge this is the first published work to provide a real-market test of an option-based indicator for asset price bubbles.

The outline of this paper is as follows.  In the following section the mathematical setting will be specified and we briefly review the martingale theory of asset price bubbles. In Section 3, we study the presence of asset price bubbles under SABR stock price dynamics and derive the corresponding martingale defect indicator. Section 4 is devoted to the introduction of the statistical framework and in Section 5 we present examples that validate the feasibility of the proposed technique using real market data related to the recent investing craze in tech stocks. We conclude with a discussion of our findings in Section 6.

\section{Strict local martingale models}
\subsection{Definition of a bubble}
This section introduces the mathematical setting and briefly reviews some key facts from the martingale theory of asset price bubbles. 

We consider a continuous time model on a finite time horizon $[0,T]$. Let $(\Omega,\mathcal{F},(\mathcal{F}_t)_{t\geq 0},\mathbb{Q})$ denote a filtered probability space, where $\mathcal{F}$ denotes the $\sigma$-field of measurable subsets of $\Omega$ and the filtration $(\mathcal{F}_t)_{t\geq 0}$ satisfies the \lq\lq usual conditions\rq\rq,\ see Protter \cite{Protter0}. $\mathbb{Q}$ is an equivalent local martingale  measure defined as follows: If the market is complete, then the equivalent local martingale measure is unique by the second fundamental theorem of asset pricing, see Jarrow and Protter \cite{Jarrow0}. If the market is incomplete, there exists a set of equivalent local martingale measures and we assume in this work that the model under consideration is embedded into a larger complete market such that the equivalent local martingale measure is unique. That is, for instance in the presence of two risk factors, reflecting market and volatility risk, we assume that in addition to the tradable strategies of holding the risky asset and a money market account, options on the risky asset can also be used for hedging purposes. In this case, the choice of the unique equivalent martingale measure is fully determined by the market price of volatility risk, which reflects the risk preferences of the market participants.

Let us consider a risky asset with stock price process $X$ starting at $X_0=x$ which has continuous paths for $\mathbb{Q}$-almost all $\omega\in\Omega$. We assume that the risk-free rate $r\geq 0$ is constant and that the stock pays a continuous dividend yield $q$ which includes the borrow cost. 
\begin{defn}\label{def:1}\normalfont
The stock price is said to \emph{admit a bubble} on $[0,T]$ with respect to the measure $\mathbb{Q}$ if the discounted value 
\begin{equation}
e^{-(r-q)t}X_t
\end{equation}
of an account that initially purchases one share of stock and continuously reinvests the dividends in the stock is a strict local martingale on $[0,T]$ with respect to $\mathbb{Q}$.
\end{defn}
The standard financial economic theory says that a bubble appears when the \emph{market value} of an asset exceeds its \emph{fundamental value}. Therefore, let us define the \emph{fundamental value} at $T$ as the discounted expected future value of $X$ 
\begin{equation}\label{eqn:ct}
m_x(T):=e^{-rT}\mathbb{E}_xX_T
\end{equation}
as well as its normalized \emph{martingale defect}
\begin{equation}\label{eqn:defect}
d_x(T):=1-e^{qT}x^{-1}m_x(T).
\end{equation}
Clearly, $e^{-(r-q)\cdot}X_{\cdot}$ is a true martingale on the interval $[0,T]$ if and only if $d_x(T)=0$ and by the local martingale property, $e^{qT}x^{-1}m_x(T)$ is $\mathbb{Q}$-a.s. bounded by $1$ and decreasing as a function of $T$. In other words, we have $d_x(T)\geq 0$ with equality if and only if there is no bubble. A strictly positive normalized martingale defect implies the presence of a bubble whose severity is quantified by the magnitude of $d_x(T)$. In Section \ref{sec:SABR} we are going to exploit this observation in order to define our so-called \emph{martingale defect indicator}.

\begin{ex}\label{ex:1}\normalfont
Consider the \emph{constant elasticity of variance (CEV)} model with elasticity parameter $\beta\geq 0$. We assume, for simplicity, that $r=q\equiv 0$, then the stock price process is given by the unique strong solution to the stochastic differential equation
\begin{equation*}
\dd X_t=X_t^{\beta}\dd W_t,\quad X_0=x.
\end{equation*}
If $\beta<1$, zero is accessible and the absorbing boundary condition terminates the trajectories once the boundary is hit. Note that for $\beta<1/2$ one could as well impose a reflecting boundary condition.  
For the choice $\beta>1$, zero is inaccessible, paths being pushed away from the origin. This case was first studied in the work \cite{Emanuel} and is the most interesting one in the context of this work as it yields an asset price bubble. For the case $\beta=2$, one can compute the density of $X$ explicitly to obtain
\begin{equation*}
\mathbb{E}_xX_T=x\left(1-2\Phi\left((-x^{\beta}\sqrt{T})^{-1}\right)\right)<x,
\end{equation*}
where $\Phi$ is the distribution function of the standard normal distribution. 
\end{ex}
\subsection{Option pricing in the presence of a bubble}
There exists different concepts in the literature for defining option prices in the presence of asset price bubbles. Let us start  by defining the price of a put, respectively call option with strike $K$ and maturity $T$ by the risk-neutral expectations of their future payoffs:
\begin{equation}\label{eqn:PutPayoff}
\mathcal{P}_x(K,T)=e^{-rT}\mathbb{E}_x(K-X_T)^+,\quad \mathcal{C}_x(K,T)=e^{-rT}\mathbb{E}_x(X_T-K)^+.
\end{equation}
It is well-known that in the presence of asset price bubbles this risk-neutral valuation bears several subtleties such as the call price not necessarily being convex as a function of the spot price and put-call parity failing in its usual form, cf., e.g., \cite{CoxHobson}. More precisely, it holds for all $(K,T)\in (0,\infty)^2$ and every $x>0$ that
\begin{equation}\label{eqn:putcall}
\mathcal{C}_x(K,T)-\mathcal{P}_x(K,T)=m_x(T)-e^{-rT}K
\end{equation}
together with the following inequalities: 
\begin{eqnarray}\label{eqn:static}
(m_x(T)-e^{-rT}K)^+&\leq& \mathcal{C}_x(K,T)<m_x(T)\\
\label{eqn:static1}
(e^{-rT}K-m_x(T))^+&\leq& \mathcal{P}_x(K,T)<e^{-rT}K.
\end{eqnarray}
Note that in the presence of an asset price bubble, equation (\ref{eqn:putcall}) implies that the usual put-call parity relation is violated. Moreover, by inequality (\ref{eqn:static}), there exists a spot price $x$ such that the lower no-static-arbitrage bound for the call price is violated.
\begin{ex}
\label{ex:2}\normalfont
We continue within the setting of Example \ref{ex:1} with elasticity parameter $\beta=2$. Note that $\mathcal{C}_x(0,T)=\mathbb{E}_xX_T$ is the price of a zero-strike call option with maturity $T$, and $\partial^2_{xx}\mathcal{C}_x<0$, that is, $\mathcal{C}_x$ is strictly concave as a function of the spot price $x$. Moreover, this example also illustrates the necessity of \lq\lq correcting\rq\rq\ the classical initial boundary value problems for the Black-Scholes-Merton backwards PDE, respectively the Dupire forward PDE for call options in the presence of a bubble because in their standard form, uniqueness would be lost, $\mathcal{C}_x(0,T):=x$ being another solution. 

More precisely, for a general one-factor local volatility model the call price $\mathcal{C}_x(K,T)$ is given by the unique bounded classical solution to the initial value problem for the modified Dupire forward equation
\begin{equation*}
\partial_T \mathcal{C}_x=\frac{\sigma^2(K)}{2}K^2\partial^2_{KK}\mathcal{C}_x-(r-q)K\partial_K\mathcal{C}_x-q(\mathcal{C}_x-m_x(T))+\partial_T m_x(T)
\end{equation*}
for all $(K,T)\in (0,\infty)^2$ with initial value $\mathcal{C}_x(K,0)=(x-K)^+$ and boundary condition
$\mathcal{C}_x(0,T)=m_x(T)$, see \cite{Tysk2}. Note that while $m_x(T)=e^{-qT}x$ yields the standard Dupire forward equation, the case $d_x(T)>0$ produces some extra terms whereas the corresponding equation for put prices remains unchanged compared to the true martingale setting. That is, put prices cannot \lq\lq see\rq\rq\ the presence of a bubble due to the bounded payoff.
\end{ex}
\subsection{Collateral requirements and bubbles}\label{subsec:collateral}
Violations of the put-call parity as described above are rarely observed in real option markets so that risk-neutral valuation is not able to match market prices in the presence of asset price bubbles. This apparent shortcoming can be resolved by accounting for collateral requirements which was first shown by Cox and Hobson \cite{CoxHobson}. They employ the concurrent concept to define option prices in the presence of bubbles via their super-replication cost, that is, one defines the option price as the smallest initial fortune required to super-replicate the corresponding payoff structure, see \cite{CoxHobson}. Cox and Hobson point out that because of short positions the super-replicating portfolios of call options have values which are unbounded from below and therefore not admissible in the sense of Delbean and Schachermayer \cite{Delbean1}. In banking practice, these admissibility restrictions are reflected in the need to deposit collateral due to margin requirements for short positions. That is, on top of replicating the payoff at maturity, the hedging portfolio for a call option is required to satisfy a collateral requirement at all times before maturity. Namely, for each $t\in [0,T]$ it should be worth at least 
\begin{equation}\label{eqn:coll1}
(e^{-q(T-t)}X_t-e^{-r(T-t)}K)^+. 
\end{equation}
Note that this requirement is automatically satisfied in the absence of bubbles due to the fact that the underlying process is a bounded local martingale thus implying that (\ref{eqn:static}) reduces to the well-known no-static-arbitrage bound. In a strict local martingale setting, by \cite[Thm. 3.4]{CoxHobson} the presence of a bubble implies that $\limsup_{n\rightarrow\infty} n\mathbb{Q}(\sup_{t\in [0,T]} X_t> n)>0.$ That is, there is a probability of order $1/n$ that the underlying will rise to value $n$, hence the hedging portfolio is not automatically worth at least the value given by (\ref{eqn:coll1}); this has to be enforced by a modification of the super-replicating portfolio. It is shown in \cite[Cor 5.3]{CoxHobson} that the smallest initial fortune of a super-replicating portfolio for a fully collateralized call is given by 
\begin{equation}\label{eqn:coll}
\mathcal{C}_x^{\text{coll}}(K,T)=\mathcal{C}_x(K,T)+(xe^{-qT}-m_x(T)).
\end{equation}

Clearly, as a consequence of $(\ref{eqn:putcall})$, the fully collateralized call price coincides with the price which is obtained by employing the standard put-call parity relation for given put prices. That is, using the fully collateralized call price (\ref{eqn:coll}) instead of the risk-neutral one restores both the put-call parity and the standard no-static-arbitrage bounds. 
Note moreover, that the price (\ref{eqn:coll}) of the fully collateralized call option coincides with the call price for strict local martingale models proposed by Madan and Yor \cite{MadanYor}.

It is remarkable and highly relevant for risk management and pricing purposes that in the presence of an asset price bubble risk-neutral valuation is not a sensible choice to price call options. In this case, pricing via super-replication is more suitable and the super-replication price of a collateralized call option differs from the price of an uncollateralized one. As the payoff of a put option is bounded, no collateral requirement of the form (\ref{eqn:coll1}) on the hedging portfolio is required and thus the prices with and without collateralization coincide. 

\begin{rem}\normalfont
Note that a market including collateral requirements such that the uncollateralized call option is not a tradable asset is compatible with Merton's no-dominance assumption \cite{Merton}. That is, there is no trading strategy whose cash flows and liquidation value are always greater than or equal to the liquidation value of the risky asset and strictly greater with a positive probability and whose cost is less than the market price of the risky asset. 
\end{rem}

\clearpage
\section{Presence of bubbles under SABR dynamics}\label{sec:SABR}

The \lq\lq stochastic alpha, beta, rho\rq\rq\ or short SABR model is an extension of the CEV model which has been introduced by Hagan, Kumar, Lesniewski and Woodward \cite{Hagan}.  It is a two-factor stochastic volatility model with the following dynamics for the stock price process under the equivalent martingale measure $\mathbb{Q}$
\begin{eqnarray}\label{eqn:SABR1}
\dd X_t &=& (r-q)X_t \dd t+\alpha_t e^{-(1-\beta)(r-q)t}X_t^{\beta}\dd W_t^{(1)},\quad X_0 = x,\\
\label{eqn:SABR2}\dd \alpha_t &=& \nu\alpha_t\dd W_t^{(2)},\quad \alpha_0 = \alpha,
\end{eqnarray}
with the elasticity parameter $\beta\in[0,1]$, the volatility of the volatility $\nu>0$ and two correlated Brownian motions $W^{(1)}$ and $W^{(2)}$ such that 
\begin{equation*}
\dd \big<W^{(1)},W^{(2)}\big>_t=\rho\dd t
\end{equation*}
for some constant $\rho\in[-1,1]$. We are mostly interested in the lognormal SABR model, i.e., $\beta=1$, in which the forward 
\begin{equation}\label{eqn:SABR3}
  F_t = 
  X_te^{(r-q)(T-t)}
\end{equation} 
for fixed maturity $T$ and for time $t \in (0,T]$ has a lognormal distribution 
\begin{equation*}
  \dd \log F_t =-\frac{1}{2}\alpha_t^2\dd t+\alpha_t\dd W_t^{(1)}.
\end{equation*}

\begin{rem}\normalfont
Of course, we are aware of the main shortcomings of the SABR model for pricing equity derivatives, its 
non mean reverting volatility and insufficient calibration to market quotes for very short dated options. We would like to emphasize that the performance of SABR in our application context is not affected at all by these shortcomings. Indeed, it is common market practice to employ SABR for inter- and extrapolation of implied volatilities which is exactly the way we are going to use it in this work. As was pointed out by Jacquier and Keller-Ressel in \cite{Jacquier}, every option-based test for asset price bubbles comes down to extrapolating the implied volatility surface. We think that SABR is an ideal choice to do so because it provides a decent fit for maturities ranging from 2 months to 3 years while at the same time we have analytical expressions for all the relevant quantities as we will deduce from Theorem \ref{thm:1} below. 
\end{rem}

Our approach is based on the following inverse calibration problem: \vspace{.25cm}

\noindent \emph{Under SABR dynamics (\ref{eqn:SABR1}), (\ref{eqn:SABR2}) with elasticity parameter $\beta=1$, specify a unique equivalent local martingale pricing measure. Given observed bid and ask put and call market quotes, find SABR parameters $\boldsymbol{\theta}=(\alpha,\nu,\rho)^T$ such that the model prices are compatible with these market quotes. }

\vspace{.25cm}
\noindent Once we have found a solution to this problem, we can determine the normalized SABR martingale defect:
\begin{equation}\label{eqn:indicator}
d_x(T;\boldsymbol{\theta})=1-{x}^{-1}e^{-(r-q)T}\mathbb{E}_x\{F_T\}. 
\end{equation}
Note, however, that the inverse calibration problem is ill-posed as it might not have a solution or solutions might not be unique.  
\begin{rem}\normalfont
As the dynamic of the SABR forward (\ref{eqn:SABR3}) and the corresponding numerics are particularly simple, we prefer working with the forward rather than the spot. 
\end{rem}

With regard to the existence of asset price bubbles in this setting, we have the following result. 
\begin{thm}\label{thm:1}
A stock price following the SABR dynamics (\ref{eqn:SABR1}), (\ref{eqn:SABR2}) with elasticity parameter $\beta=1$ and parameters 
$\boldsymbol{\theta}=(\alpha,\nu,\rho)^T$ admits a bubble on $[0,T]$ if and only if $\rho>0$. 
Moreover, the normalized SABR martingale defect
is given by
\[
  d_x(T;\boldsymbol{\theta})  = 1 - \exp(-2\rho\alpha/\nu) -
  A_c(\rho\alpha/\nu,\nu^2T),
\]
where
\[
  A_c (\gamma,\tau) =
  \sqrt{\frac{2\gamma}{\pi^3}} e^{-(\gamma + \mbox{$\frac18$} \tau)}
  \int_0^\infty \frac{8s\sinh(\pi s)}{4s^2 +
  1} K_{is}(\gamma) \exp(-\mbox{$\frac12$}s^2 \tau) \dd s,
\]
  and $K_\mu(\cdot)$ is the modified Bessel function of the second
  kind. The normalized SABR martingale defect is rapidly decreasing for small
  $T$, i.e., $d_x(T;\boldsymbol{\theta}) \le cT^n$ for every $n > 0$ and for large $T > 0$ the normalized SABR martingale defect satisfies the
  asymptotic expansion
  \[
    d_x(T;\boldsymbol{\theta}) = 1 - \exp(-2\gamma) -
    \frac{8\sqrt{\gamma}}{\sqrt{\nu^6T^3}} e^{-(\gamma + \mbox{$\frac18$}
    \nu^2T)}
    K_0(\gamma)(1 + \mathcal O(T^{-\frac12})).
  \]
  
\end{thm}
The mathematical proof of Theorem \ref{thm:1} is provided in Appendix \ref{app:1}. It is a direct consequence of this result that under lognormal SABR dynamics for given SABR parameters $\boldsymbol{\theta}=(\alpha,\nu,\rho)^T$  the \emph{martingale defect indicator} 
\begin{equation}\label{eqn:defect_indicator}
A(\boldsymbol{\theta}):= \lim_{T\rightarrow\infty}d_x(T;\boldsymbol{\theta})
\end{equation} 
can be computed analytically in a particularly simple form, namely 
\begin{equation}
A(\boldsymbol{\theta})= 1 - \exp(-2\rho\alpha/\nu).
\end{equation} 
In addition, a high order asymptotic expansion of the implied volatility function $\sigma_{\text{imp}}(\cdot;\boldsymbol{\theta})$ in terms of the forward (\ref{eqn:SABR3}) is readily available for the lognormal SABR model , cf. \cite{Hagan}. That is, there is no need to solve a pricing PDE neither for solving the inverse calibration problem, nor for the computation of the martingale defect indicator (\ref{eqn:defect_indicator}) which is particularly appealing in a statistical inverse problems framework where these steps have to be repeated many thousand times.

\section{Statistical indicator and estimation algorithm}\label{Sec:Indicator}

We adopt a statistical perspective on the problem of computing the martingale defect indicator: All quantities are considered as random variables so that the solution to the statistical inverse problem is the posterior probability distribution of the martingale defect indicator conditioned on the observed bid and ask market quotes. In contrast to deterministic methods, our approach incorporates the available prior knowledge in a fully explicit way. Rather than a single point estimate it can produce possibly very different estimates, all compatible with this prior knowledge, and quantify their uncertainty. 

Let us assume that the stochastic model for the stock price dynamics can be parameterized by $n$ parameters, i.e., $n=3$ for  SABR dynamics. Let $(\Omega',\mathcal{G},\mathbb{P})$ denote a probability space and let 
\begin{equation}
(\boldsymbol{\Theta},\Eb):\Omega'\rightarrow\mathbb{R}^{n+k},\quad \boldsymbol{Y}:\Omega'\rightarrow \mathbb{R}^k
\end{equation}
denote random vectors on this probability space. We use capital letters for random vectors and lower case letters for their realizations.  
The vector $(\boldsymbol{\Theta},\boldsymbol{E})$ represents the quantities that cannot be directly observed, i.e., in our setting the unknown parameters $\boldsymbol{\theta}= (\alpha,\nu,\rho)^T$ whose values control the quality of the calibration of the model to the observed market data and the unknown observation error $\boldsymbol{E}$. $\boldsymbol{Y}$ represents the vector of observable quantities, i.e., the corresponding model-implied volatilities. Those random variables are connected via the forward model 
\begin{equation}
\boldsymbol{Y}=F(\boldsymbol{\Theta},\boldsymbol{E})
\end{equation}
where the operator $F$ yields the implied volatilities $\sigma_{\text{imp}}(K_i;\boldsymbol{\theta})$ for the option prices 
$\mathcal{P}_x(K_i,T;\boldsymbol{\theta})$, resp. $\mathcal{C}_x(K_i,T;\boldsymbol{\theta})$, $i=1,...,k$, corresponding to the realization $\boldsymbol{\theta}$ of the parameter vector $\boldsymbol{\Theta}$.
These implied volatilities are computed for the time to maturity $T$ and $k$ different strikes and we assume that they are polluted by observation errors corresponding to the realization $\e\in\mathbb{R}^k$ of the error vector $\Eb$. The probability distribution of the random vector $\boldsymbol{Y}$ conditioned on the vectors $\boldsymbol{\theta}$ and $\boldsymbol{e}$ is given by
\begin{equation}
 \pi(\boldsymbol{y}|\boldsymbol{\theta},\e)=\delta(\boldsymbol{y}-F(\boldsymbol{\theta},\e)),
\end{equation}
where $\delta$ denotes Dirac's delta in $\mathbb{R}^k$. If $\pi_{\text{pr}}$ denotes the prior probability density of $(\boldsymbol{\Theta},\boldsymbol{E})$, then we may write the joint probability density of $(\boldsymbol{\Theta},\Eb)$ and $\boldsymbol{Y}$ as
\begin{equation}\label{eqn:den}
\pi(\boldsymbol{\theta},\e,\boldsymbol{y})=\pi(\boldsymbol{y}|\boldsymbol{\theta},\e)\pi_{\text{pr}}(\boldsymbol{\theta},\e)=\delta(\boldsymbol{y}-F(\boldsymbol{\theta},\e))\pi_{\text{pr}}(\boldsymbol{\theta},\e). 
\end{equation}
For simplicity we assume here that $\boldsymbol{\Theta}$ and $\Eb$ are independent random variables and that the observation noise is additive, i.e., $F(\boldsymbol{\Theta},\Eb)=f(\boldsymbol{\Theta})+\Eb$. Then we obtain from (\ref{eqn:den}) by integration that
\begin{equation*}
\pi(\boldsymbol{\theta},\boldsymbol{y})=\pi_{\text{pr}}(\boldsymbol{\theta})\pi_{\text{noise}}(\boldsymbol{y}-f(\boldsymbol{\theta})). 
\end{equation*}
In particular we can formulate the following statistical inverse calibration problem: 

\vspace{.25cm}
\noindent\emph{Compute the posterior distribution of $\boldsymbol{\Theta}$ conditioned on the observed market data $\boldsymbol{y}$ which is given by Bayes' formula
\begin{equation}\label{eqn:Bayes}
\pi(\boldsymbol{\theta}|\boldsymbol{y})=\frac{\pi(\boldsymbol{\theta},\boldsymbol{y})}{\int_{\mathbb{R}^n} \pi(\boldsymbol{\theta},\boldsymbol{y})\dd \boldsymbol{\theta}}.
\end{equation}}

\noindent Given the solution to this statistical inverse problem, we can compute the posterior density for our quantity of interest, the martingale defect indicator 
\begin{equation}
\pi(A(\boldsymbol{\theta})|\boldsymbol{y})
\end{equation}
together with a variety of estimates as well as a posteriori uncertainty measures for these estimates, see, e.g., Kaipio and Somersalo \cite{KaSo}.

The unnormalized posterior density reads
\begin{equation}
\pi(\boldsymbol \theta \vert \boldsymbol y) \propto \pi(\boldsymbol\theta) \pi(\boldsymbol y \vert \boldsymbol\theta),
\end{equation}
where  $\pi(\boldsymbol \theta)$ and $\pi(\boldsymbol y \vert \boldsymbol\theta)$ are the prior and likelihood probability density, respectively.
We factorize the prior as 
\begin{equation}
\pi(\boldsymbol\theta)  = \pi(\alpha)\pi(\nu)\pi(\rho) =  [\alpha\in \mathbb{R}][\nu \geq 0] [ \vert \rho\vert \leq 1 ],
\end{equation}
where we have a flat prior for $\alpha$, flat prior in $\mathbb{R}^+$ for $\nu$, and a uniform prior for $\rho\in [-1,1]$ and for notational convenience, we have used the Iverson bracket $[\cdot ]$ as an indicator function:
\begin{equation*}
[B] := \begin{cases} 1 , & \textrm{if $B$ is true,} \\ 0, & \textrm{otherwise.} \end{cases}
\end{equation*}
We note that our prior construction is an improper prior, but in practical numerical computations in connection with the likelihood density, the posterior density becomes a proper probability density. This means that although sampling  from an improper prior density is not feasible, sampling the posterior is nevertheless feasible. For a discussion on using improper priors in MCMC sampling schemes, see Hobert and Casella 1996 \cite{Hobert1996}.

We assume that the observation error is Gaussian such that the likelihood function is given by
\begin{equation}
\pi(\boldsymbol y \vert \boldsymbol\theta) \propto \exp\left(-\frac{1}{2}  ( \boldsymbol{y}-f(\boldsymbol\theta))^T \Sigma^{-1}( \boldsymbol{y}-f(\boldsymbol\theta))\right)\prod_{i=1}^k \left[ \vert  \boldsymbol{y}_i-f_i(\boldsymbol\theta) \vert \leq \frac{1}{2}\text{BA}_i\right],
\end{equation}
where $\Sigma$ is  covariance matrix of the observation error $\boldsymbol E$, and $\text{BA}_i$ is the bid-ask spread at the $i$-th strike in terms of the implied volatilities. The corresponding posterior proves to be difficult to study analytically, as we have a non-linear parameter estimation problem with somewhat complex priors and constraints. 

Two standard numerical techniques to cope with this complexity are to use optimization or Markov chain Monte Carlo (MCMC) methods. 
Although optimization methods are computationally fast, they can merely provide point estimates rather than uncertainty quantification through sampling of the posterior. MCMC sampling methods on the other hand are commonly used to enable uncertainty quantification in parameter estimation problems. We shall consider adaptive single-component Metropolis-Hastings based on, e.g.,\ Haario et al.\ \cite{Haario} and Roberts and Rosenthal \cite{Roberts}. However, MCMC methods are known to be very sensitive with respect to initial values, hence, we start in step 1 by searching for good initial values for $\rho$ and $\nu$ using optimization. More precisely, we compute a MAP estimate by maximizing the posterior density using the Nelder-Mead algorithm. As the convergence of such a local search method depends on the choice of a sufficiently good initial guess, we compute such an initial guess using the method described by West in \cite{West}. Subsequently in Step 2, we run adaptive MCMC for all three parameters. For a more detailed description, see Algorithm \ref{alg:Optim_and_adaptiveMCMC} below. 
\begin{algorithm}[!h]
	\algsetup{linenosize=\small}
	\small
	\caption{Combined optimization and adaptive MCMC}\label{alg:Optim_and_adaptiveMCMC}
	\begin{algorithmic}[1]	
		\REQUIRE Initial guesses \(\rho_0\) and \(\nu_0 \)
		\STATE  Step 1. Set $\beta = 1$.	
 Estimate $\rho$ and $\nu$ via Nelder-Mead algorithm with initial values \(\rho_0\), \(\nu_0 \) and at each iteration step,
 find $\alpha$ as a cubic root as described in \cite{West}.
 		\STATE Step 2. Estimate $\rho$, $\nu$, and $\alpha$ via MCMC using $J$ samples and initial values obtained from optimization in Step 1.

		\FOR{$j=1$ to $J$}	\vspace{1mm}
			\STATE Draw $\alpha \vert \alpha^{(j-1)} \sim \mathcal{N}\left(\alpha^{(j-1)} ,s_{\alpha}^2\right)$
			\STATE Compute $p_{\alpha}=\min\left\{1, \frac{\pi\left(\alpha^{(j-1)} \vert \rho^{(j-1)},\nu^{(j-1)} \right) \pi(\alpha)}{\pi\left(\alpha \vert  \rho^{(j-1)},\nu^{(j-1)} \right) \pi\left(\alpha^{(j-1)}\right)}\right\}.$ 
			\STATE With probability $p_{\alpha}$ set $\alpha^{(j)}=\alpha$, otherwise set $\alpha^{(j)}=\alpha^{(j-1)}$.
			\STATE Run Adaptation for $s_{\alpha}^2$.
			\STATE Draw $\rho \vert \rho^{(j-1)} \sim \mathcal{N}\left(\rho^{(j-1)} ,s_{\rho}^2\right)$
			\STATE Compute $p_{\rho}=\min\left\{1, \frac{\pi\left(\rho^{(j-1)}\right) \vert \alpha^{(t)},\nu^{(j-1)}) \pi(\rho)}{\pi\left(\rho \vert  \alpha^{(j)},\nu^{(j-1)}\right) \pi\left(\rho^{(j-1)}\right)}\right\}.$ 
			\STATE With probability $p_{\rho}$ set $\rho^{(j)}=\rho$, otherwise set $\rho^{(j)}=\rho^{(j-1)}$.
			\STATE Run Adaptation for $s_{\rho}^2$.
			\STATE Draw $\nu \vert \nu^{(j-1)} \sim \mathcal{N}\left(\nu^{(j-1)} ,s_{\nu}^2\right)$
			\STATE Compute $p_{\nu}=\min\left\{1, \frac{\pi\left(\nu^{(j-1)} \vert \alpha^{(j)},\rho^{(j)}\right) \pi(\nu)}{\pi\left(\nu \vert \alpha^{(j)},\rho^{(j)}\right) \pi\left(\nu^{(j-1)}\right)}\right\}.$ 
			\STATE With probability $p_{\nu}$ set $\nu^{(j)}=\nu$, otherwise set $\nu^{(j)}=\nu^{(j-1)}$.
			\STATE Run Adaptation for $s_{\nu}^2$.
			\STATE Update martingale defect indicator $A^{(j)}(\boldsymbol{\theta}^{(j)})= 1 - \exp\left(-2\rho^{(j)}\alpha^{(j)}/\nu^{(j)}\right)$.
		\ENDFOR
	\end{algorithmic}
	 \label{algorithm:MCMC}
\end{algorithm}
\clearpage

\section{Real-market examples}

There are market concerns that compare the current tech stock rally to the period of excessive speculation from 1997 to 2001 that was based on the internet hype and resulted in the so-called \lq\lq dot-com bubble\rq\rq\ . In fact, tech sector indices such as the Nasdaq Composite or the S\&P 500 Information Technology Index have already surpassed former peak levels that were set during that period. From a historical point of view, tech stocks seem to be expensive both in relation to earnings and in comparison to other sectors. Therefore the tech sector should provide a promising data set for real-market tests for the statistical indicator presented in this work. 

With regard to tech companies, in particular for those with fast growth and high uncertainty, there is a demand for complementary approaches to validate stock price movements. The massive volatility observed in this setting reflect both the complexity when it comes to valuation issues and the limitations of fundamental analysis when it comes to evaluate the company's intangible assets and to forecast reliable longterm cash flows. Shorthand metrics traditionally used to determine whether a stock is overvalued, such as price-earnings or price-to-book multiples, are of limited use in the context of negative earnings and book values. The following real-market examples highlight how our option-based indicator could generate added value in this setting.

\subsection{Market data}
We use composite Nasdaq option chain market data for SNAP Inc. (ISIN US83304A1060)\footnote{https://www.nasdaq.com/symbol/snap/option-chain}, for Twitter Inc. (ISIN US90184L1026) \footnote{https://www.nasdaq.com/symbol/twtr/option-chain} and for Square Inc. (ISIN US8522341036) \footnote{https://www.nasdaq.com/symbol/sq/option-chain}. 
We consider only one single maturity $T$.
First, to enhance informative value we filter this data from a liquidity perspective as follows:
\begin{itemize}
\item Eliminate all quotes with no volume
\item Discard all quotes from in-the money options
\item Eliminate all quotes $< 0.03$ USD 
\item Discard quotes in order to obtain the longest monotonic subsequence of mid prices (not necessarily unique).
\end{itemize}
Now we estimate the market implied forward, which depends on the risk-free interest rate, the dividend yield and the borrow cost. While the risk free interest rate (we use the USD Swap OIS Fed Funds rate) and the short-term dividend-yield are readily available, the borrow cost has to be implied from observed market prices. If the borrow cost is correctly identified, implied volatilities for put and call options with identical strike and maturity should more or less coincide. We start with an initial guess $q^{\text{est}}$ for the implied dividend yield including borrow cost and proceed as follows:
\begin{enumerate}
\item[(i)] Compute mid market prices from call and put bid and ask quotes.
\item[(ii)] Compute mid implied volatilities using the current estimate $q^{\text{est}}$ for the implied dividend yield: As the exercise style of market quoted single stock options is American, we use the bisection method, a Crank-Nicolson finite difference discretization and the projected SOR method to solve the corresponding free boundary value problem for the option prices. 
\item[(iii)] Use the current estimate $q^{\text{est}}$ together with the implied volatilities from (ii) to compute for the nearest strikes $K^{\text{l}}$, $K^{\text{u}}$ on either side of the current estimate of the forward 
the corresponding European option prices $\mathcal{P}_x^{\text{e}}(K^{\text{l,u}},T)$, respectively $\mathcal{C}_x^{\text{e}}(K^{\text{l,u}},T)$.  Compute the new estimate $F^{\text{est}}$ for the forward with maturity $T$ as the average of the put-call parity implied forwards $F^{\text{l}}$ and $F^{\text{u}}$:
\begin{equation*}
F^{\text{l,u}}=e^{rT}\cdot (\mathcal{C}_x^{\text{e}}(K^{\text{l,u}},T)-\mathcal{P}_x^{\text{e}}(K^{\text{l,u}},T))+K^{\text{l,u}}.
\end{equation*}
\item[(iv)] Compute the new estimate for the implied dividend yield including the borrow cost via
\begin{equation*}
q^{\text{est}} = \frac{1}{T}\log\left(\frac{xe^{rT}}{F^{\text{est}}}\right).
\end{equation*}
\end{enumerate}
We iterate this until the put and call implied mid volatilities agree within $0.1\%$. Then we use the resulting implied dividend yield including the borrow cost to compute the implied bid and ask volatilities from the market quotes as described in (ii).
\subsection{Are tech stocks in a bubble in 2018?}
\subsubsection{SNAP Inc.}
\begin{figure}[t!]
\centering
{\includegraphics[width=11cm]{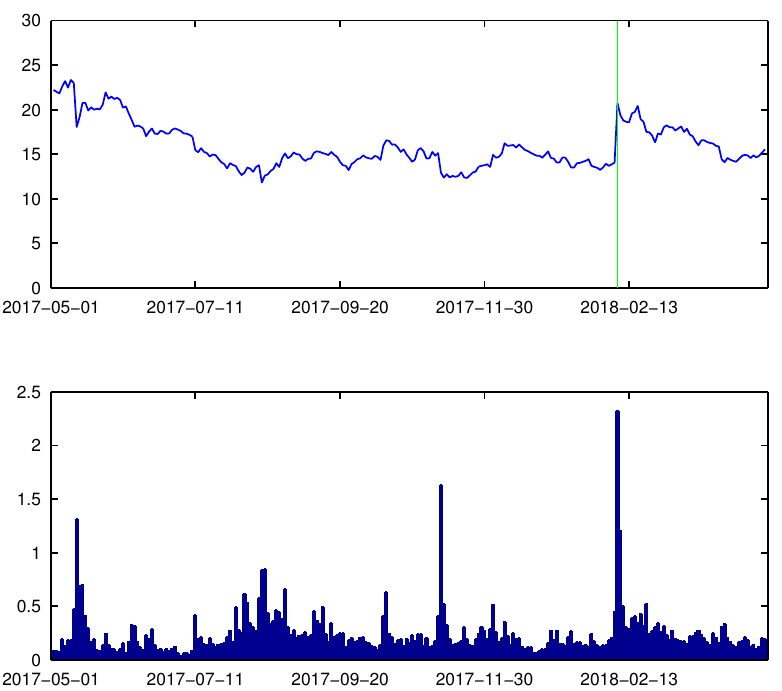}}
\caption{SNAP Inc. stock prices (USD) and trading volume (100 M)}\label{fig:SNAP_one_year}
\end{figure}

\begin{figure}[t!]
\centering
{\includegraphics[width=10.9cm]{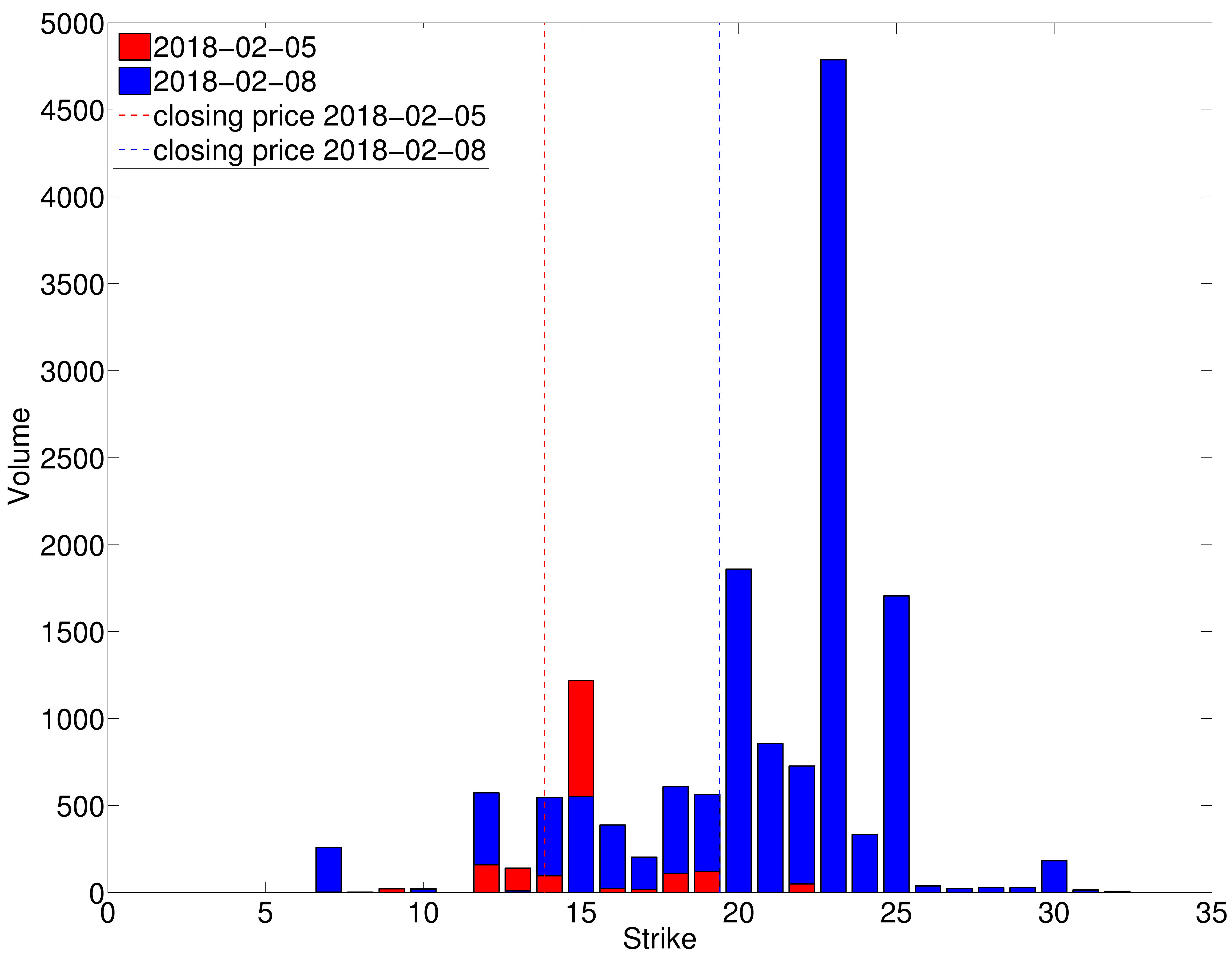}}
\caption{Volume of call options on SNAP Inc. with maturity April 20th 2018}\label{fig:SNAP_volume_calls}
\end{figure}

\begin{figure}[b!]
\centering
{\includegraphics[width=10.9cm]{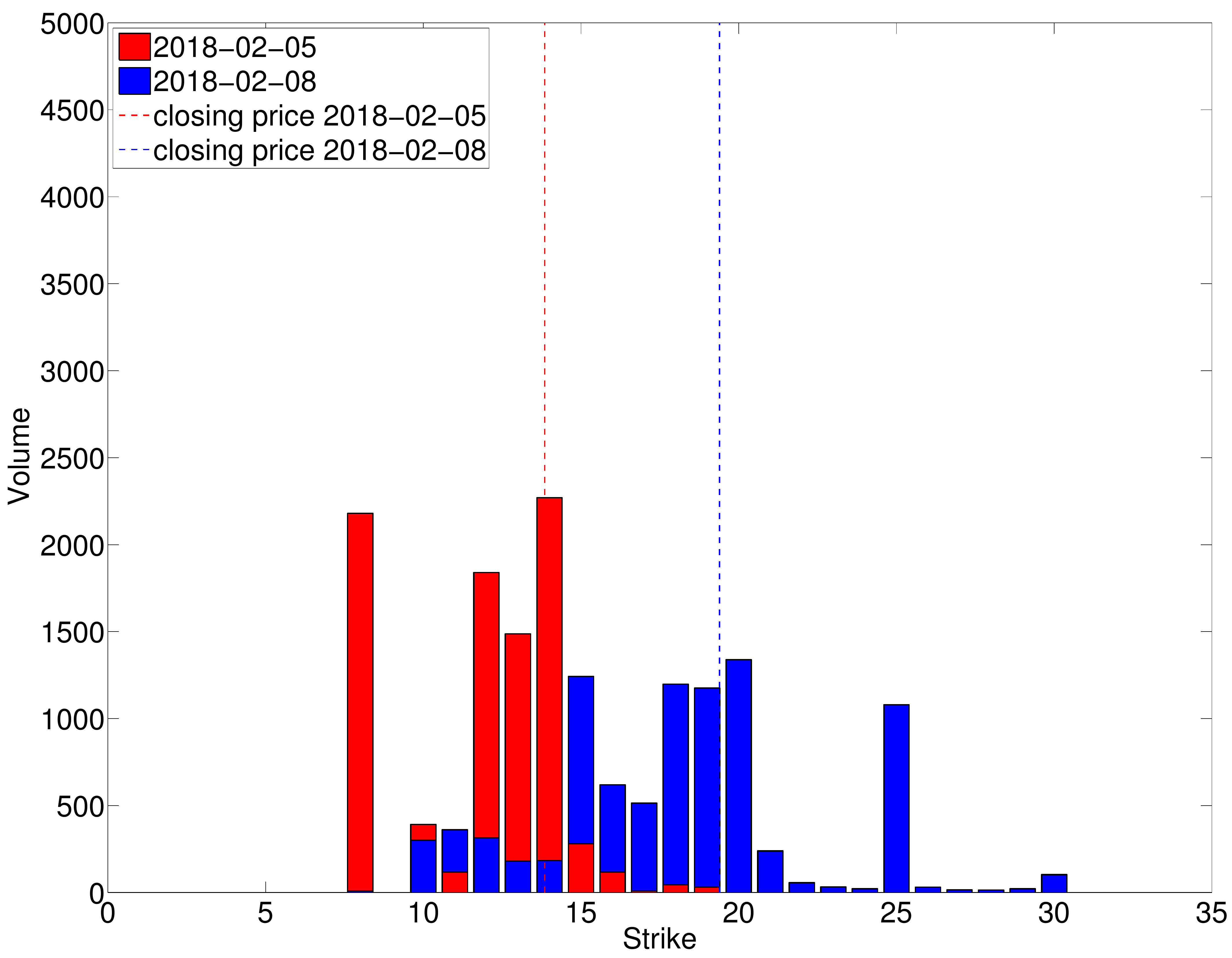}}
\caption{Volume of put options on SNAP Inc. with maturity April 20th 2018}\label{fig:SNAP_volume_puts}
\end{figure}

Following the initial hype of going public there was a massive decline in valuation of SNAP Inc. accompanied by downside peaks in response to disappointing quarterly announcements. With a $48\%$ gain in one single day, SNAP Inc.'s 4Q17 earnings release after the closing bell on February 6th 2018 triggered a highly unusual one-day share price move. SNAP Inc. shares skyrocketed after surprisingly beating analysts' expectations for the first time since its stock market launch a year ago. Historical closing price movements and the corresponding trading volume for SNAP Inc. are shown in Figure \ref{fig:SNAP_one_year}. Note the remarkable peak in the trading volume after the 4Q17 earnings release. This was only to a certain degree fueled by positive sentiment generated by sell-side analyst reports following the announcement. In fact, short interest in SNAP Inc.'s stock had reached all-time high levels by the end of 2017 so that it is plausible to assume that short covering has had a considerable influence on the stock price movement. Investors'\ behavior can be illustrated with reference to the derivatives markets. We have compared the data as of February 5th and February 8th, see Figures \ref{fig:SNAP_volume_calls} and \ref{fig:SNAP_volume_puts}, respectively. Here a sharp jump in the volume of exchange traded call options occurred while the trading volume of put options did not change dramatically. The dominant volume of put option contracts on February 5th reflects a risk-averse behavior of market participants that switched from hedging to speculative purposes reflected by the increase in volume of both near-the-money and out-of-the money calls on February 8th. This persisting high volume in put options suggests that although traders covered some of their short exposure in fear of a continuing rally the net sentiment in the stock was still bearish.
 
\begin{figure}[t!]
\centering
{\includegraphics[width=11cm]{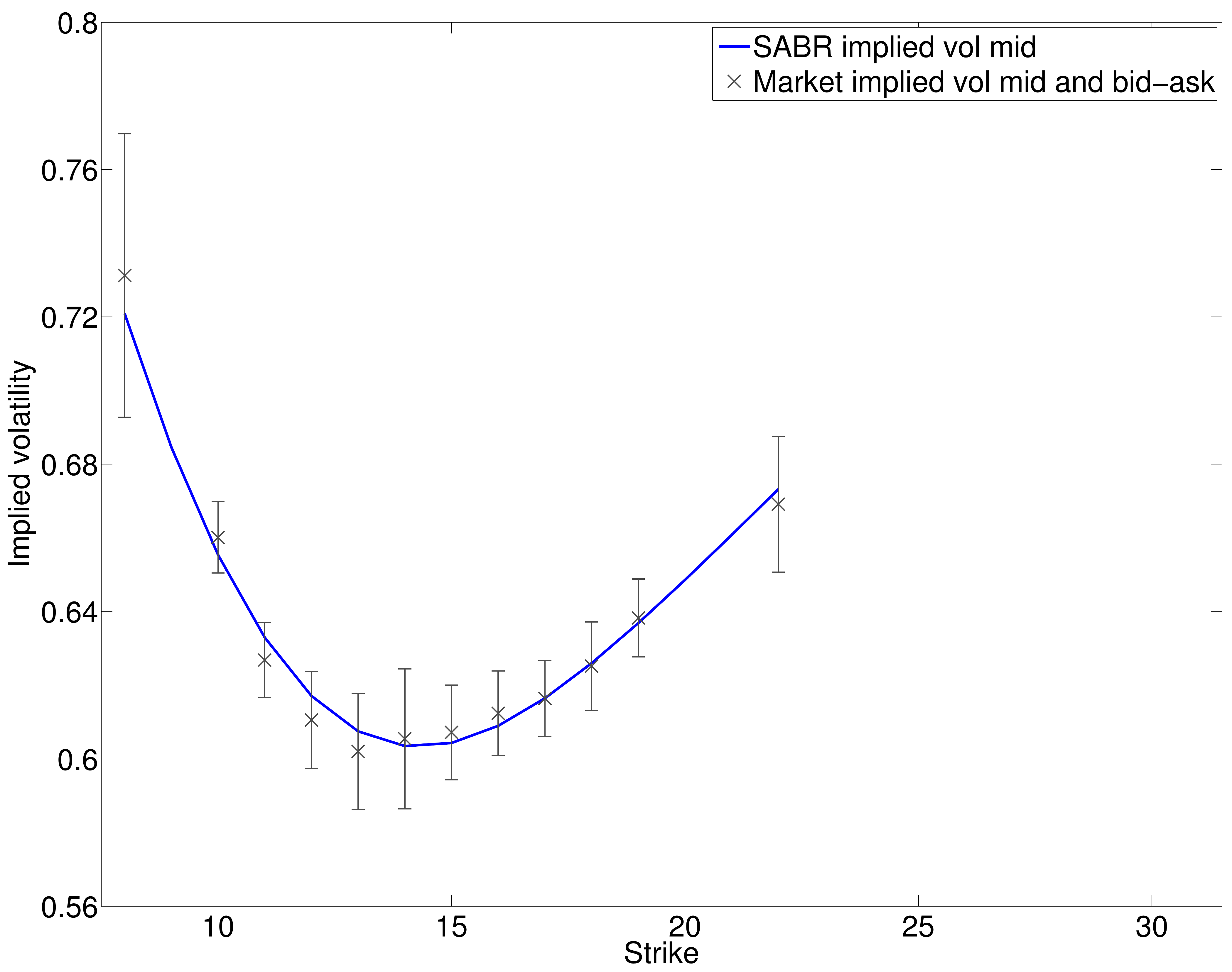}} 
\caption{SNAP Inc. February 5th 2018\ -- Market and SABR implied volatility.}\label{fig:SNAP05FEB_vol}
\end{figure}

\begin{figure}[b!]
\centering
{\includegraphics[width=11cm]{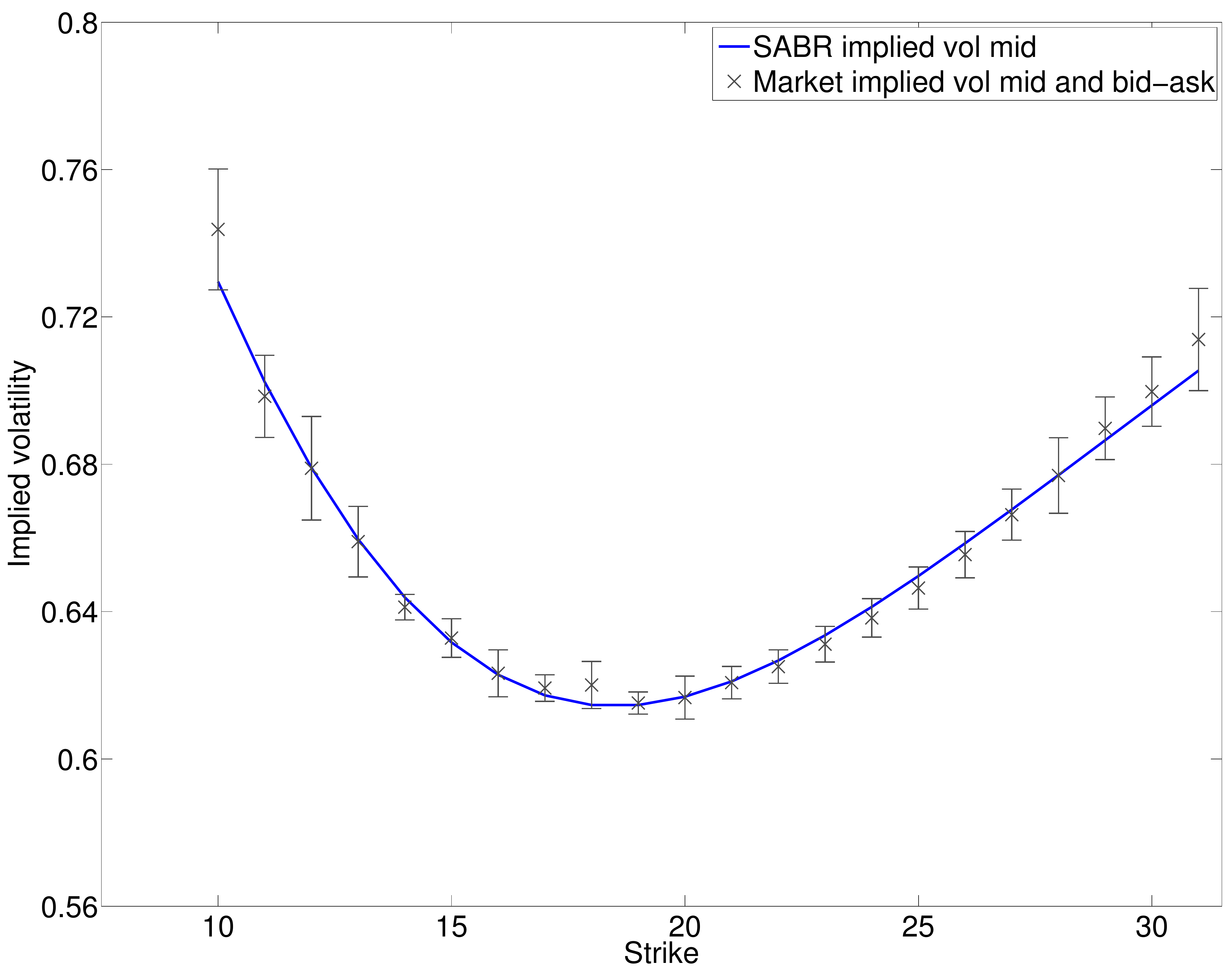}} 
\caption{SNAP Inc. February 8th 2018\ -- Market and SABR implied volatility.}\label{fig:SNAP08FEB_vol}
\end{figure}

\begin{figure}[t!]
\centering
{\includegraphics[width=11cm]{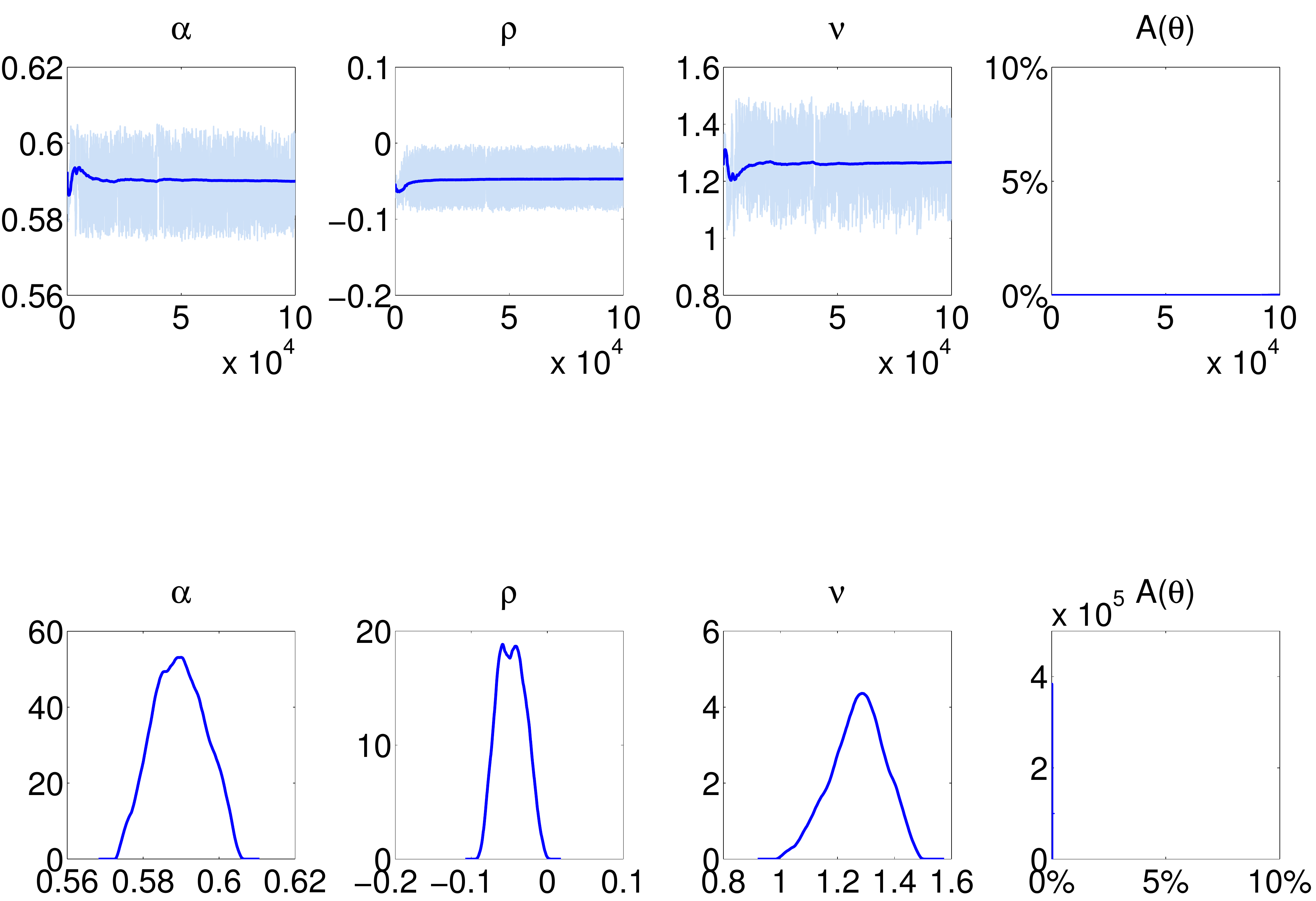}} \vspace{.2cm}
\caption{SNAP Inc.\ Feb 5 2018\ -- MCMC chains for $\alpha,\rho,\nu$ and martingale defect indicator $A(\boldsymbol{\theta})$ with corresponding marginal densities in the bottom row.}
\label{fig:SNAP05Feb}
\end{figure}

\begin{figure}[b!]
\centering
{\includegraphics[width=11cm]{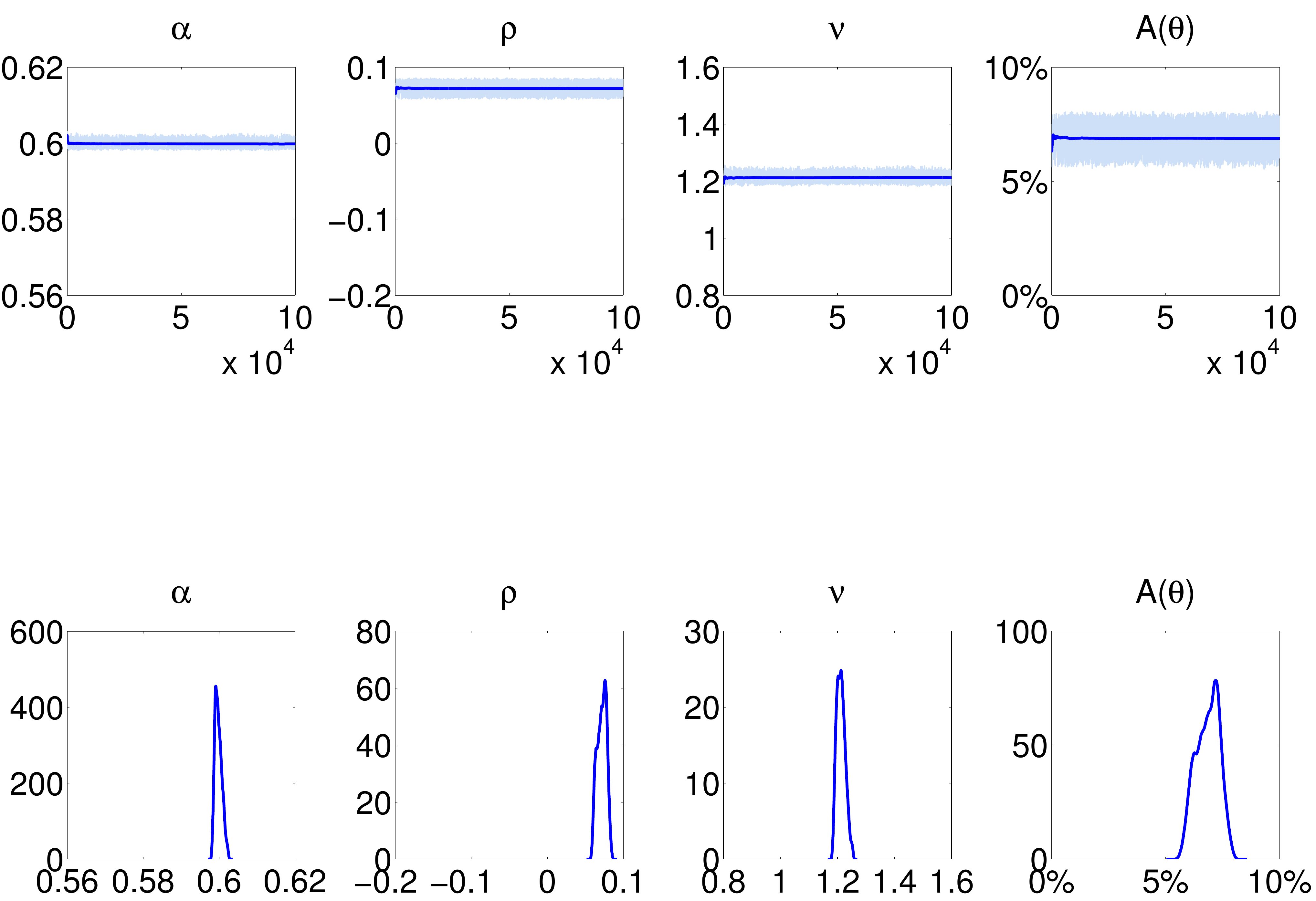}} \vspace{.2cm}
\caption{SNAP Inc.\ Feb 8 2018\ -- MCMC chains for $\alpha,\rho,\nu$ and martingale defect indicator $A(\boldsymbol{\theta})$ with corresponding marginal densities in the bottom row.}
\label{fig:SNAP08Feb}
\end{figure}

We have solved the statistical inverse calibration problem for both dates, February 5th and February 8th and computed the corresponding martingale defect indicators. We have used the option chains with maturity April 20th 2018. For the observation error we have assumed zero-mean white noise with unit variance. Note that we could add a more elaborate noise model here, but for the sake of simplicity, we leave the question of how to sensibly model observation noise for future work. The SABR implied volatilities based on the conditional mean of the sampled posterior densities are plotted in Figures \ref{fig:SNAP05FEB_vol} and \ref{fig:SNAP08FEB_vol}, respectively. Obviously, the increase in liquidity for the exchange traded options has lead to tighter bid-ask spreads and on top of that on February 8th there is liquidity for a broader range of out-of-the-money strikes. Note the slight W-shaped form of the implied volatility in Figure \ref{fig:SNAP08FEB_vol} indicating that near-the-money puts are particularly expensive, a nowadays quite common phenomenon around earnings dates of tech stocks. In order to enable a SABR fit we have artificially doubled the bid-ask-spread for the puts with strike $18$ USD and we justify this by the fact that the observed bid ask-spreads are extremely tight imposing hard constraints on the calibration. Except for this outlier, it can be said that both SABR fits are very good and compatible with the observed market quotes. In Figures \ref{fig:SNAP05Feb} and \ref{fig:SNAP08Feb}, we show traceplots and plots of cumulative averages for  the estimates of $\alpha,\rho,\nu$ and the martingale defect indicator $A(\boldsymbol{\theta})$. We also plot the marginal densities obtained by removing the burn-in period (25\% per cent of the whole chain length), and by using kernel density estimator with Epanechnikov kernels with bandwidth $\vert \max(\alpha^{(j)})-\min(\alpha^{(j)})\vert/15$, where $\max(\alpha^{(j)})$ and $\min(\alpha^{(j)})$ are the maximum and minimum values of the chain $\alpha^{(j)}$, $j=1,...,J$. We use analogous bandwidths for the other parameters.
By visually assessing, we note that we have good mixing of the chains, and we don't need to use excessively long MCMC runs, i.e.,\ here the chain length is 100,000, and this is quite enough. However, we note that without the optimization part described in Algorithm \ref{algorithm:MCMC}, the chains would not converge. As can be seen from Figure \ref{fig:SNAP05Feb}, the wide bid-ask spreads and limited amount of available market data on February 5th allow for quite a wide range of parameter values $\boldsymbol{\theta}$ which are all compatible with the observed market quotes. However, none of these parameter values yields a strict local martingale setting with $A(\boldsymbol{\theta})>0$. Based on this analysis we can be confident that there was no stock price bubble of the type defined in Definition \ref{def:1}. \pagebreak

\noindent The situation is different for February 8th, as can be seen from Figure \ref{fig:SNAP08Feb}. We have used the February 5th scales for the plot of the chains in order to visualize how the increase in liquidity reduces uncertainty in the statistical inverse problem. All of the sampled parameter values lead to strictly positive values of the martingale defect indicator and its conditional mean is $6.85\%$. This apparent bubble was most probably propelled by margin requirements for short sellers. Our statistical indicator does not detect any bubbles on any of the following days. Indeed, the stock lost nearly $20\%$ until April 2nd thus correcting the exuberance.
\FloatBarrier

\FloatBarrier

\FloatBarrier

\FloatBarrier

\FloatBarrier

\FloatBarrier

\FloatBarrier

\subsubsection{Twitter Inc.}
\begin{figure}[b!]
\centering
{\includegraphics[width=12cm]{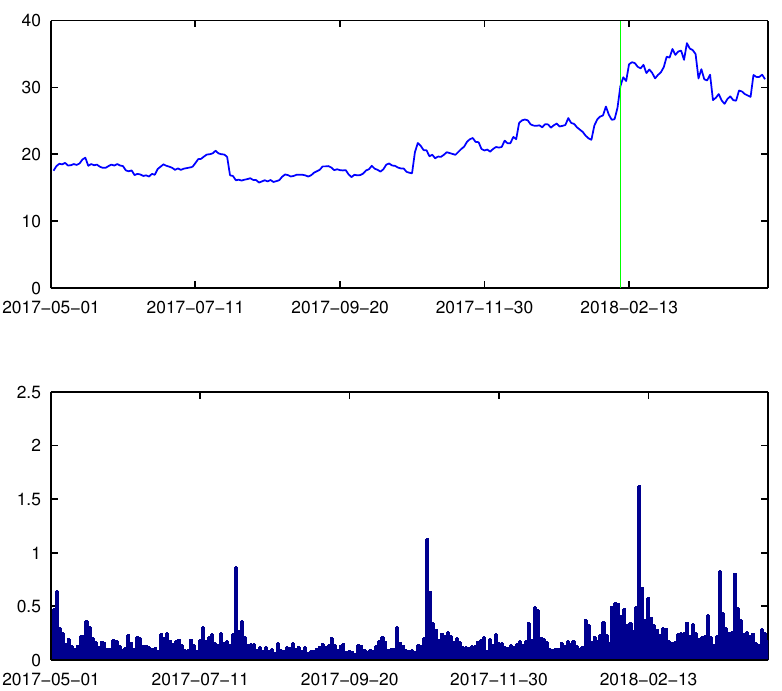}}
\caption{Twitter Inc. stock prices (USD) and trading volume (100 M)}\label{fig:TWTR_one_year}
\end{figure}

\begin{figure}[t!]
\centering
{\includegraphics[width=10.9cm]{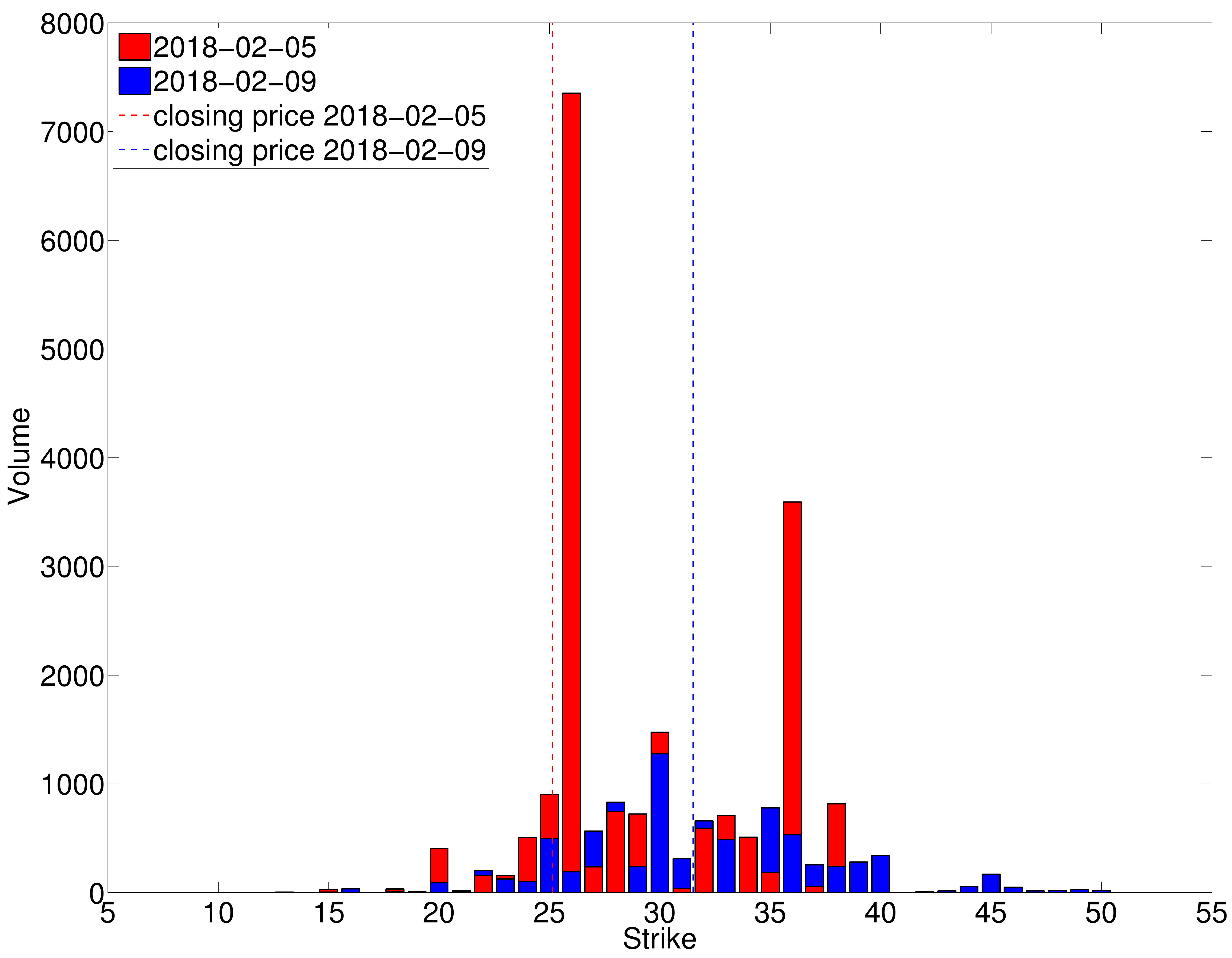}}
\caption{Volume of call options on Twitter Inc. with maturity June 15th 2018}\label{fig:TWTR_volume_calls}
\end{figure}

\begin{figure}[b!]
\centering
{\includegraphics[width=10.9cm]{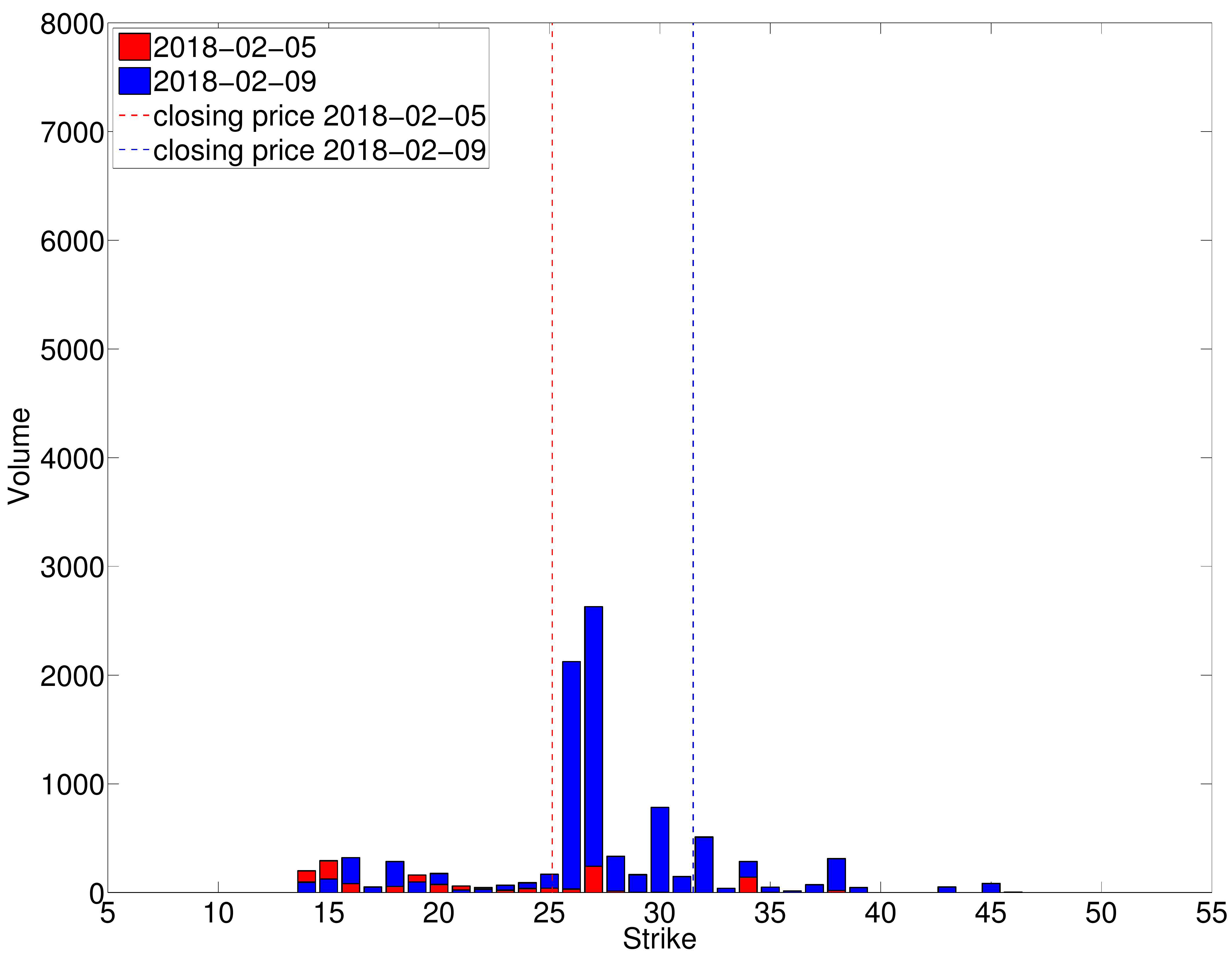}}
\caption{Volume of put options on Twitter Inc. with maturity June 15th 2018}\label{fig:TWTR_volume_puts}
\end{figure}

\begin{figure}[t!]
\centering
{\includegraphics[width=10.9cm]{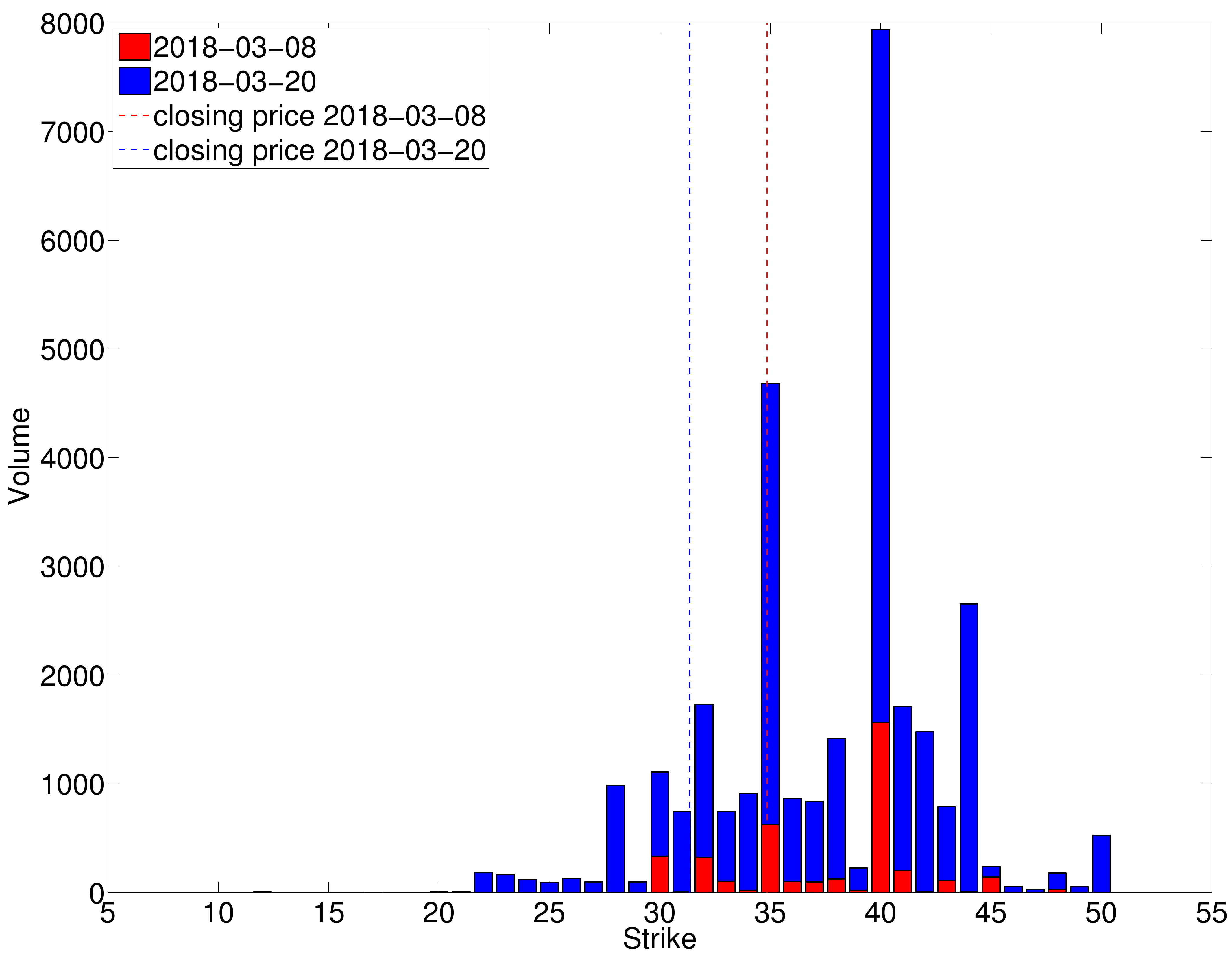}}
\caption{Volume of call options on Twitter Inc. with maturity June 15th 2018}\label{fig:TWTR_volume_calls_1}
\end{figure}

\begin{figure}[b!]
\centering
{\includegraphics[width=10.9cm]{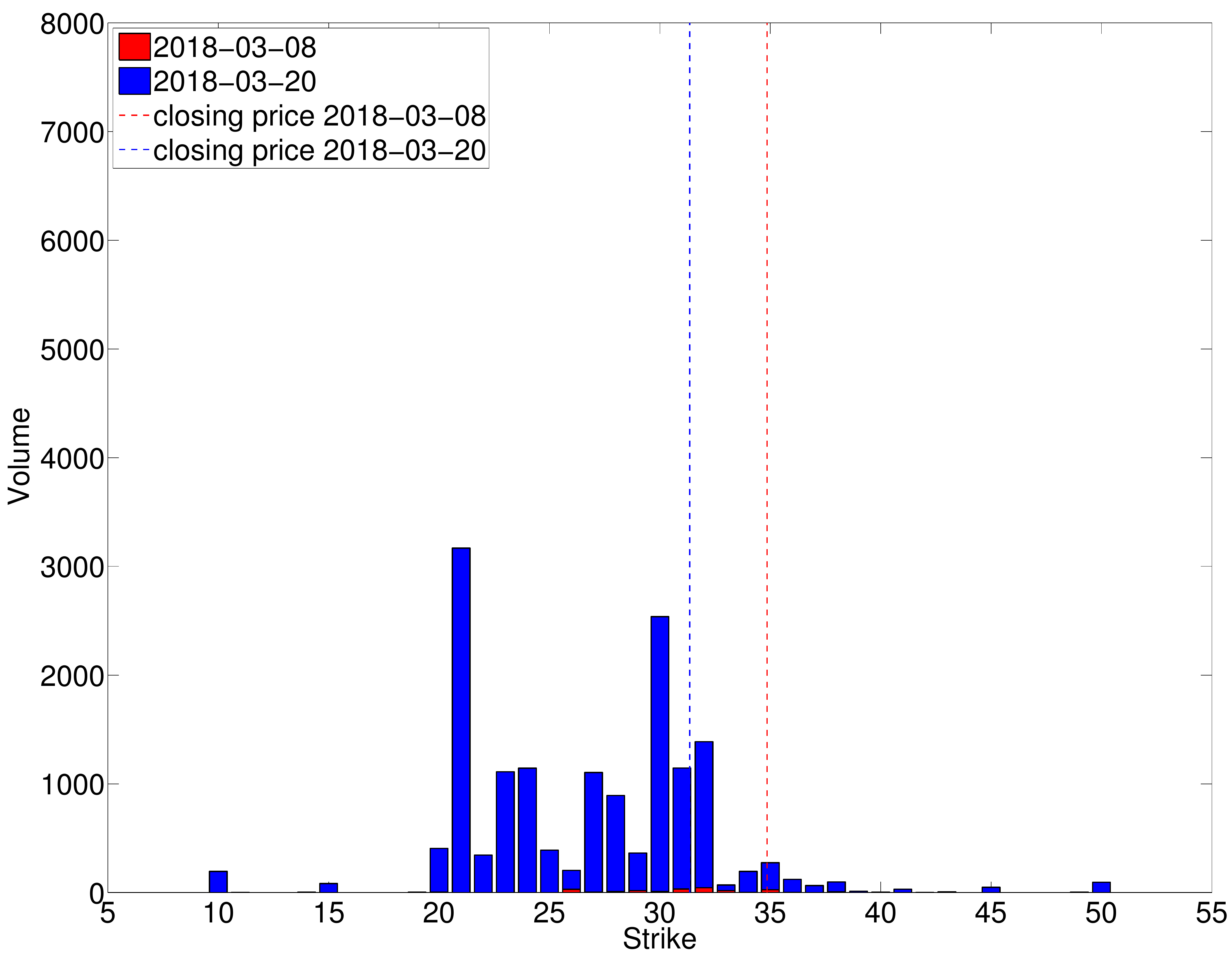}}
\caption{Volume of put options on Twitter Inc. with maturity June 15th 2018}\label{fig:TWTR_volume_puts_1}
\end{figure}

Before the opening bell on February 8th 2018 Twitter Inc. delivered the first profitable quarter in its 12-year history and reported its return to revenue growth. As the financial results beat Wall Street analysts' expectations the social network's shares jumped more than 20 percent during the subsequent trading session. Historical closing price movements and the corresponding trading volume for Twitter Inc. are shown in Figure \ref{fig:TWTR_one_year}. Note the remarkable peak in the trading volume after the 4Q17 earnings release. In order to illuminate the situation before and after the earnings release, we have compared the call and put options volume as of February 5th and February 9th, see Figures \ref{fig:TWTR_volume_calls} and \ref{fig:TWTR_volume_puts}, respectively. While the put volume increased sharply, there was at the same time a slight decrease in the call volume. The dominant volume of put option contracts on February 9th reflects a risk-averse behavior of market participants after the jump due to the earnings surprise. Indeed, our statistical indicator does not detect any bubble on both days and we have omitted the corresponding plots for the sake of brevity. However, we have found the first sign of a stock price bubble on February 14th when suddenly many market participants switched from hedging to speculative behavior. According to our indicator, this bubble persisted until March 19th 2018.
As an example, we have compared the call and put options volume as of March 8th and March 20th, see Figures \ref{fig:TWTR_volume_calls_1} and \ref{fig:TWTR_volume_puts_1}, respectively. The bullish sentiment is reflected in the pronounced volume of both near-the-money and out-of-the money calls and the vanishingly small volume of put options on March 8th. In contrast to that, some market participants had obviously returned to risk aversion on March 20th which is illustrated by the increase in put option volume. On the other hand there were also many market participants taking a contrarian point of view after the setback by using near-the-money and out-of-the money calls for speculative purposes. As a background information it should be added that on March 20th the markets witnessed high volatility in the technology sector led by growing concern about data regulatory in the wake of the Facebook Cambridge Analytica scandal. 

\begin{figure}[t!]
\centering
{\includegraphics[width=11cm]{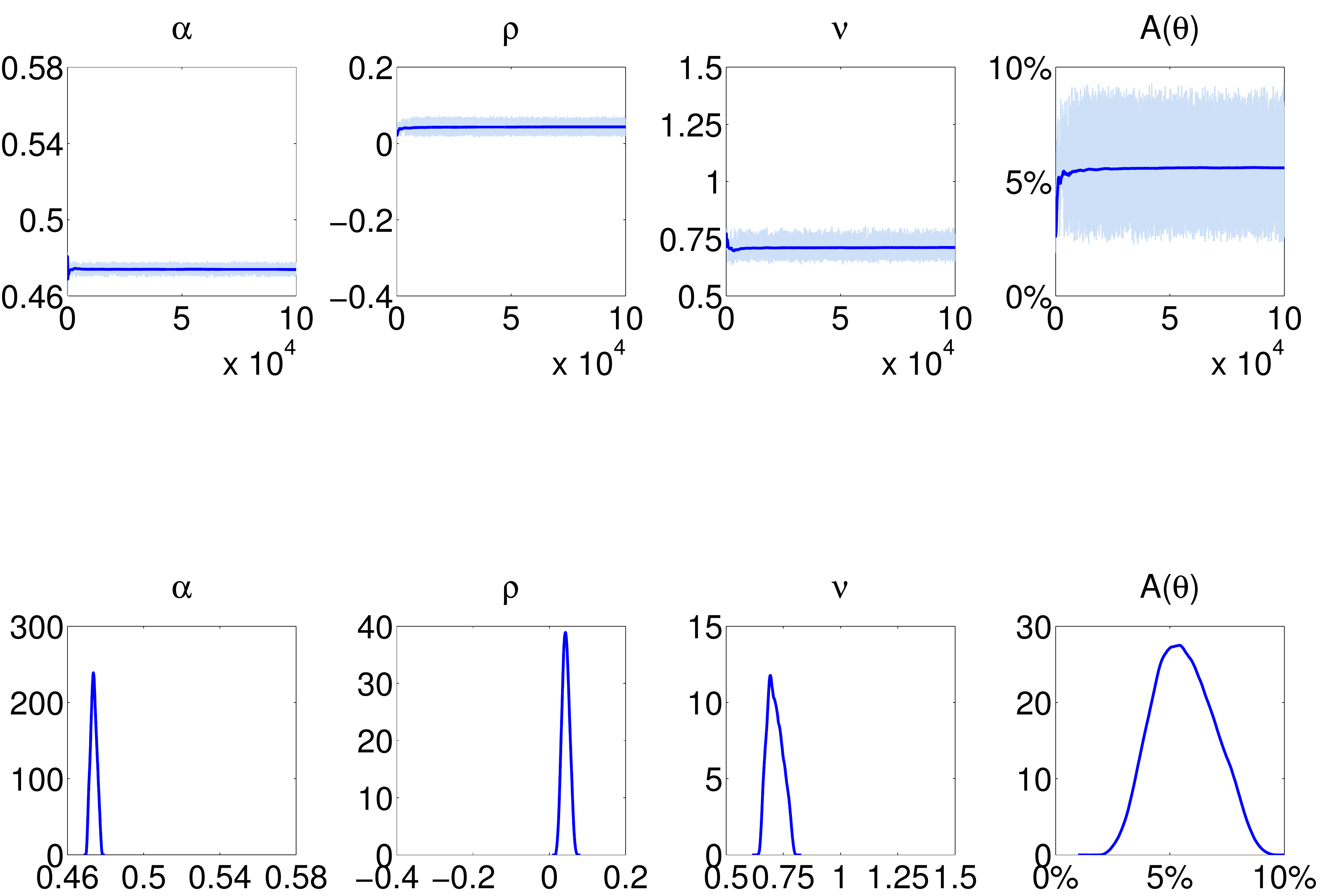}} \vspace{.2cm}
\caption{Twitter Inc.\ Feb 14 2018\ -- MCMC chains for $\alpha,\rho,\nu$ and martingale defect indicator $A(\boldsymbol{\theta})$ with corresponding marginal densities in the bottom row.}
\label{fig:TWTR14Feb}
\end{figure}

\begin{figure}[b!]
\centering
{\includegraphics[width=11cm]{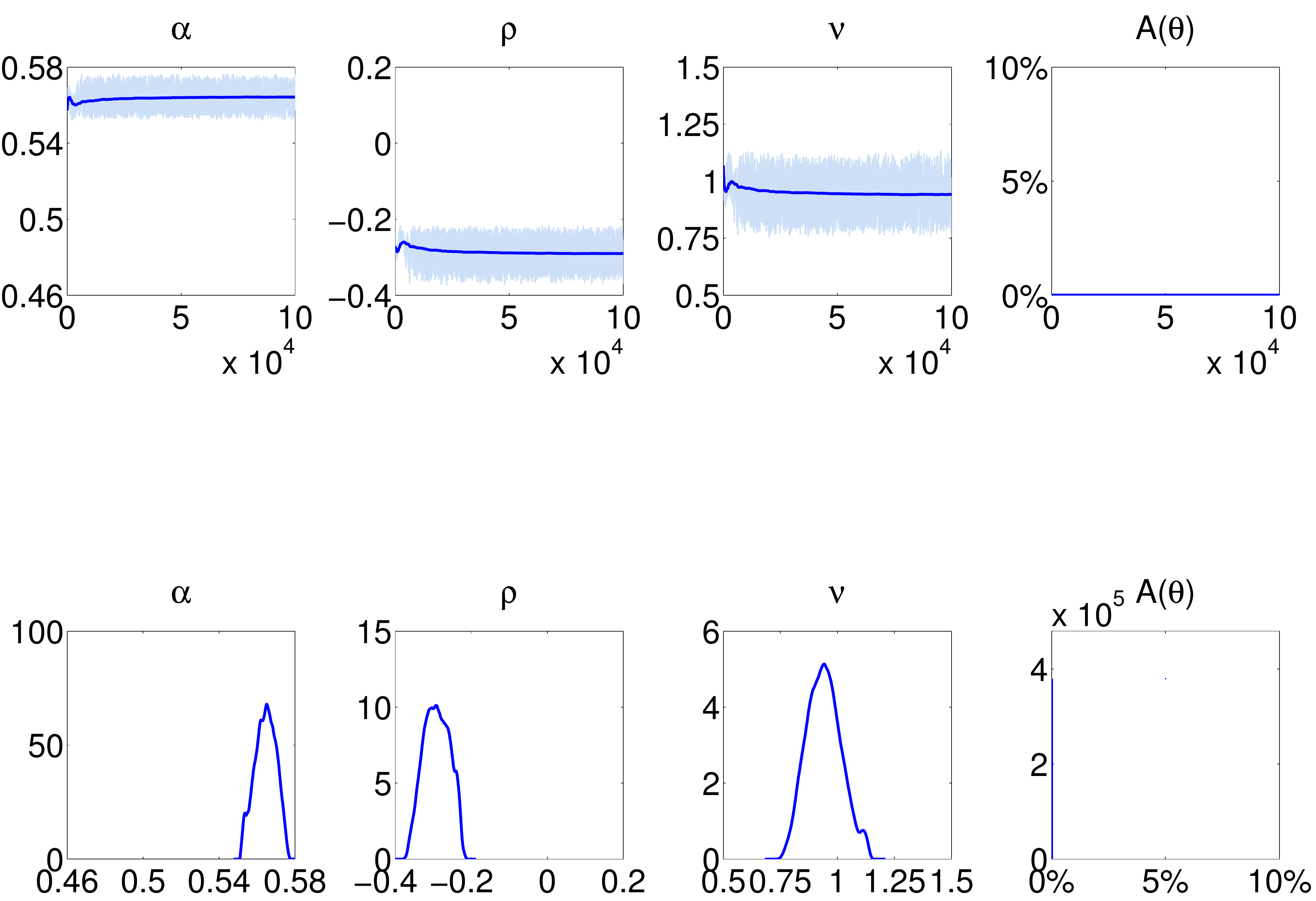}} \vspace{.2cm}
\caption{Twitter Inc.\ Mar 20 2018\ -- MCMC chains for $\alpha,\rho,\nu$ and martingale defect indicator $A(\boldsymbol{\theta})$ with corresponding marginal densities in the bottom row.}
\label{fig:TWTR20Mar}
\end{figure}

We have solved the statistical inverse calibration problem for the two dates which mark the beginning and the end of the exuberance, February 14th and March 20th and computed the corresponding martingale defect indicators. We have used the option chains with maturity June 15th 2018. For the observation error we have assumed zero-mean white noise with unit variance. In Figures \ref{fig:TWTR14Feb} and \ref{fig:TWTR20Mar}, we show traceplots and plots of cumulative averages for  the estimates of $\alpha,\rho,\nu$ and the martingale defect indicator $A(\boldsymbol{\theta})$. We also plot the marginal densities obtained by removing the burn-in period (25\% per cent of the whole chain length), and by using kernel density estimator with Epanechnikov kernels analogously to the previous example. By visually assessing, we note that we have good mixing of the chains, and we don't need to use excessively long MCMC runs, i.e.,\ here the chain length is 100,000, and this is quite enough. Again, we note that without the optimization part described in Algorithm \ref{algorithm:MCMC}, the chains would not converge. As can be seen from Figure \ref{fig:TWTR14Feb}, there is a comparably small amount of uncertainty present in the statistical inverse problem on February 14th. Tight bid-ask spreads lead to sharp peaks for the parameter marginal densities and all of these possible parameter values imply a strict local martingale setting with $A(\boldsymbol{\theta})>0$ and a conditional mean of $5.56\%$ for the martingale defect indicator. The situation is very different for March 20th, as can be seen from Figure \ref{fig:TWTR20Mar}. None of the sampled parameter values lead to strictly positive values of the martingale defect indicator. That is, this short-term stock price bubble had finally burst and our statistical indicator did not detect any bubbles on any of the following days. 

\begin{rem}\normalfont
Note that the detected short-term bubble in the stock price of Twitter Inc. is fundamentally different from the short-selling fueled short-term bubble in SNAP Inc.'s stock from the previous example. While the latter burst almost immediately, the former persisted for more than one month. In fact, the short-term bubble in Twitter Inc.'s stock price was finally deflated by an exogenous event. Therefore it is recommendable to use the proposed martingale defect indicator in combination with further company and market information.

\end{rem}

\FloatBarrier

\FloatBarrier

\FloatBarrier

\FloatBarrier

\FloatBarrier

\FloatBarrier

\newpage

\subsubsection{Square Inc.}
As a last example we compute the martingale defect indicator on June 1st for Square Inc's put and call options with maturity September 21st 2018. We provide this example without any further background information as its purpose is mainly to demonstrate the usefulness of our indicator's inherent uncertainty quantification. 

\begin{figure}[h!]
\centering
{\includegraphics[width=11cm]{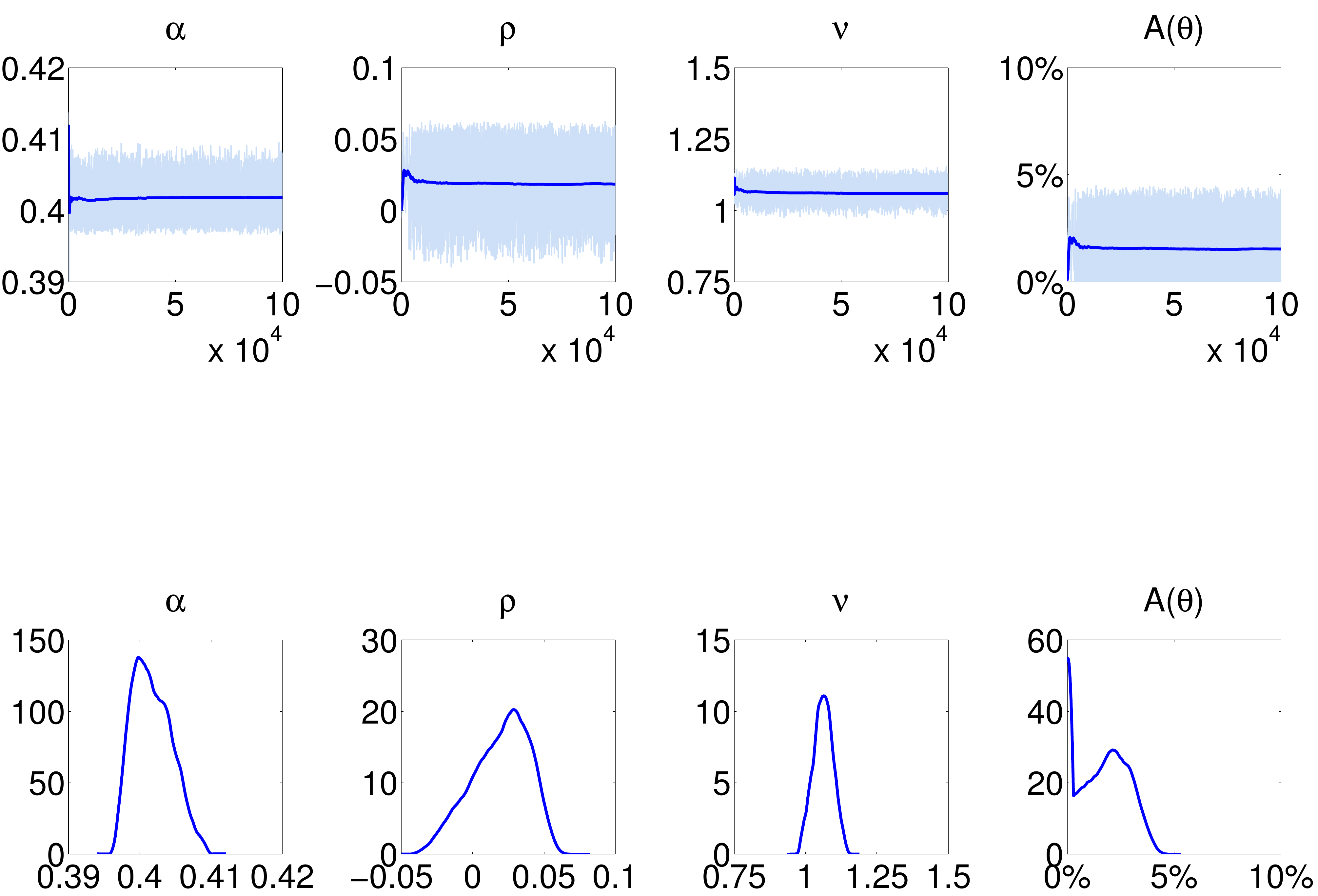}} \vspace{.2cm}
\caption{Square Inc.\ June 1st 2018\ -- MCMC chains for $\alpha,\rho,\nu$ and martingale defect indicator $A(\boldsymbol{\theta})$ with corresponding marginal densities in the bottom row.}
\label{fig:SQR01Jun}
\end{figure}

In Figure \ref{fig:SQR01Jun}, we show the usual traceplots and plots of cumulative averages for  the estimates of $\alpha,\rho,\nu$ and the martingale defect indicator $A(\boldsymbol{\theta})$. We also plot the marginal densities obtained by removing the burn-in period (25\% per cent of the whole chain length), and by using kernel density estimator with Epanechnikov kernels. As can be seen by the density of $A(\boldsymbol{\theta})$ there are basically two parameter regimens, one high and narrow peak with a significant mass in zero and another smaller but wider peak with a positive martingale defect indicator. While this uncertainty is obvious in our framework, a deterministic indicator based on, e.g., optimization can only distinguish between \lq\lq bubble\rq\rq\ or \lq\lq no bubble\rq\rq. Through uncertainty quantification, the limitation of the models, methodology and data employed can be assessed.

\FloatBarrier

\newpage

\section{Conclusion}
The ability to detect a mismatch between the market value and the fundamental value in the case of growth stocks driven by speculative market behavior is crucial for investing purposes. In this work we have constructed a statistical martingale defect indicator for detecting short-term asset price bubbles. In the provided real-market examples we focus on the tech sector due to increasing market rumors of a tech bubble 2.0. We should note here, that beyond the examples presented in this work, we have detected bubbles in other tech stocks as well. In the majority of cases these occurred after earnings announcements and persisted on a timescale between several days and one month. Eventually most of these bubbles were corrected by sharp downside peaks. It is thus one of our main experimental findings that short-term bubbles, which burst after several days or weeks, might be more common than we think. While the current work concentrates on the mathematical background and three proof-of-concept examples, we are planning to present extensive real-market tests in a follow-up paper.

With regard to the afore mentioned limitations when it comes to the validation of tech stock prices, the presented statistical indicator provides a valuable risk management tool for investors trading in the corresponding stocks. We propose to apply the indicator complementary to fundamental analysis and to consider the corresponding marginal densities for uncertainty assessment. When it comes to detecting long-term bubbles, the option-based approach seems inappropriate as the liquidity in options with maturity longer than two years is very limited and implied volatility surfaces beyond these maturities are usually flat. However, it seems plausible that in the presence of many short-term bubbles at the same time not all of them are always completely corrected by the market. The resulting mispricings might add up to one long-term market bubble in a whole sector. With regard to the tech sector this concern is supported by the fact that most of the investment crazes and stock market bubbles in history have occurred in times of economic and technological transformation. Indeed, technology sector indices have more than recovered from the implosion of the dot-com bubble and recently set new all time highs despite regulatory and protective concerns that could trigger a rally ending exogenous event. With regard to possible spillover effects it is remarkable that at the same time the S\&P 500 Index's already significant tech stock exposure was boosted by adding rebounding Twitter Inc. shares, replacing seed and agricultural chemicals maker Monsanto. 
\newpage

\appendix
\section{Proof of Theorem \ref{thm:1}}\label{app:1}
\begin{proof}
  Since for the fixed maturity $T$ the discounted value $X_t e^{-(r-q)t}$ is a constant multiple of the
  forward price $F_t$, the existence of a bubble (cf.
  Definition~\ref{def:1}) is equivalent to the
  forward price being a strict local martingale. The following argument
  is a combination of slightly modified arguments due to
  Sin~\cite{Sin98} and Lewis~\cite{Lewis98}.
  The forward $F$ and stochastic volatility $\alpha$ given by
  equations~\eqref{eqn:SABR3} and~\eqref{eqn:SABR2} are both positive exponential
  semimartingales
  \begin{equation}
    F_t = F_0 \mathcal E(\alpha \cdot W^{(1)})_t \quad\text{and}\quad
    \alpha_t = \alpha \mathcal E(\nu W^{(2)})_t,
  \end{equation}
  respectively. Here we denote the exponential semimartingale of $X$
  with $X_0 = 0$ as
  \[
    \mathcal E(X)_t = \exp\left(X_t - \frac12 \langle X \rangle_t\right)
  \]
  and the $H \cdot X$ denotes the stochastic integral with respect to a
  semimartingale 
  \[
    (H \cdot X)_t = \int_0^t H(s) \dd X_s.
  \]
  By Fatou's Theorem, the positive exponential semimartingale $\mathcal
  E(X)$ is a martingale if and only if $\EW \mathcal E(X)_t = 1$ for
  every $t > 0$. 
  As Cox and Hobson~\cite{CoxHobson} point out, the martingale property
  follows with an argument due to Sin~\cite{Sin98}.
  According to this work, when $\beta = 1$, the
  expectation of the exponential semimartingale $F_t/F_0 = \mathcal
  E(\alpha \cdot W^{(1)})$ is given by
  \[
    \EW \mathcal E(\alpha \cdot W^{(1)})_t = \widehat{\mathbb Q}(
    \widehat{\tau}_\infty > t)
  \]
  for every $t > 0$ where under the $\widehat{\mathbb Q}$-probability the stopping time
  $\widehat{\tau}_\infty$ is the time of explosion of the auxiliary
  processes that under $\widehat{\mathbb Q}$-probability satisfies the following
  SDE
  \[
    \dd v_t = \nu v_t \dd W_t^{(3)} + \nu \rho v_t^2 \dd t,\quad v_0 =
    \alpha,
  \]
  where $W^{(3)}$ is a standard Brownian motion under $\widehat{\mathbb
  Q}$.
  When $\rho = 0$, the non-explosion is
  evident, since then $v_t = \alpha \mathcal E(\nu W^{(3)})_t$. This
  implies by the Comparison Theorem~\cite{IkWa1981} for solutions of
  stochastic differential equations, that explosion cannot occur when $\rho < 0$. 
  
  Sin~\cite{Sin98} verified using the Feller test that the explosion never occurs 
  if and only if $\rho \le 0$ but he did not compute the
  martingale defect when $\rho > 0$. 
  
  In order to compute the normalized martingale defect, we need to
  determine the complementary distribution function of $\widehat
  \tau_\infty$.  The
  explosion time can be transformed to a first hitting time to zero by
  introducing an auxiliary process $\eta_t = \nu/(v_t\rho)$. Using the
  It\^{o} formula, we notice that under the $\widehat{\mathbb Q}$-probability the
  $\eta$ satisfies a SDE 
  \[
    \dd \eta_t = \nu^2( \eta_t - 1) \dd t - \nu \eta_t \dd
    W_t^{(4)}, \quad \eta_0 = \gamma^{-1}
  \]
  when $\gamma = \rho \alpha / \nu$. Moreover, with a time change $X_t =
  \eta_{t/\nu^2}$ we can simplify the equation even further and we
  obtain a SDE for the auxiliary process $X$ given by
  \begin{equation} \label{eqn:expl_SDE}
  \dd X_t = (X_t - 1) \dd t - X_t \dd W_t, \quad X_0 = \gamma^{-1}.
  \end{equation}
  If $\tau_0$ denotes the first time the process $X$ hits zero, we
  have the identity $\widehat{\tau}_\infty = \tau_0/\nu^2$, and thus,
  \[
    \widehat{\mathbb Q}(\widehat{\tau}_\infty > t) = \mathbb
    P_{\gamma^{-1}}(\tau_0 > \nu^2 t),
  \]
  where $\mathbb P_{\gamma^{-1}}$ denotes the $\widehat{\mathbb Q}$-probability given
  $X_0 = \gamma^{-1}$.
  Moreover, the normalized martingale
  defect is given by the probability of reaching zero before the
  maturity $T$, i.e.,
  \[
    d_x(T) =  \mathbb P_{\gamma^{-1}}(\tau_0 \le
    T') = \mathbb P_{\gamma^{-1}}( X_{T' \wedge \tau_0} = 0)
  \]
  where the last identity means that the stopped process $X^{\tau_0}_t =
  X_{t \wedge \tau_0}$ has stopped before the rescaled time of maturity $T'
  = \nu^2 T$. Lewis~\cite{Lewis98} derived both the Laplace transform
  for the transition probability 
  \[
    u(t,z) = \mathbb P_{2/z} (X_{2t \wedge \tau_0} = 0)
  \]
  and the explicit solution via the inverse Laplace transform. 
  Namely,
  the Laplace transform
  \[
    g(s,z) = \int_0^\infty e^{-s t} u(t,z) \dd t
  \]
  of the transition probability $u$ satisfies for fixed $s > 0$ the differential equation 
  \[
    \begin{cases}
    z^2 \partial_z^2 g(s,z) + z^2 \partial_z g(s,
      z) - s g(s,z) = 0, & z > 0,\\
      g(s, +\infty) = s^{-1}, \quad
      g(s, 0) = 0& \\
    \end{cases}
  \]
  As in Lewis~\cite{Lewis98} multiplying the Laplace transform $g$ with
  $z^{-a}$, the function $H(z) = z^{-a} g(s,z)$ satisfies the Kummer
  equation 
  \[
    z\partial_z^2 H(z) + (2a+z) \partial_z H(z) + aH(z) = 0, \quad z > 0
  \]
  if $a^2 - a - s = 0$ or if $a = a_\pm = \frac 12 \pm \sqrt{\frac14 +
  s} = \frac12 \pm \lambda$. By
  the boundary conditions and well-known properties of confluent
  hypergeometric functions (see~\cite{AbramowitzStegun, Lewis98}), the function $g$ is determined to be
  \[
    g(s,z) = \frac1s \frac{\Gamma(a_+)}{\Gamma(2a_+)} z^{a_+} M(a_+,
    2a_+, -z)
  \]
  where $M$ denotes the Kummer's confluent hypergeometric of the first kind $M(a,b,x) =
  {}_1\mkern-1mu F_1(a;b;x)$, like in the proof of the \cite[Proposition 2.1]{Lewis98}.

  In this special case, the Kummer's function can be represented with the
  modified Bessel function of the first kind, namely
  \[
    g(s,z)  
    = \frac1s \sqrt{\pi z} e^{-z/2} I_{\lambda}(z/2)
  \]
  where the last identity follows from Kummer's transformation, the
  representation form $M(a_+,2a_+,z)$ in terms of modified Bessel
  function and the duplication formula for the gamma function.

  Like in \cite[Proposition 2.1]{Lewis98}, we can invert this by
  Bromwich integral. The infinity behavior for the absorption
  probability comes from the pole $s = 0$ corresponding $\lambda = 1/2$
  and since $I_{1/2}$ has a representation in terms of elementary
  functions, the probability of reaching zero eventually is 
  \[
    \sqrt{\pi z} e^{-z/2} I_{1/2}(z/2) 
    = 1 - e^{-z} = 1 - e^{-2\gamma}.
  \]
  This can also be derived from~\cite[Proposition 2.1]{Lewis98}, which has a different
  representation in terms of the incomplete gamma function. The contour
  integration along the branch cut from $(-\infty, -\frac14)$ gives the
  contribution of not reaching zero before time of maturity
  \[
    \begin{split}
      &\frac{\sqrt {\pi z}e^{-z/2}}{2\pi i} \int_0^\infty \frac{e^{-(s+
    \frac14)t}}{s + \frac14}
      (I_{i\sqrt s}(z/2)-I_{-i\sqrt s}(z/2))\dd s\\
      &= 
      -\sqrt {2\gamma/\pi^3}e^{-\gamma} e^{-\frac18T'} \int_0^\infty
    \frac{8s \sinh(s\pi)}{4s^2 + 1}
      K_{is}(\gamma) e^{-\frac12s^2T'}\dd s\\
    \end{split}
  \]
  which can also be derived from~\cite[Proposition 2.1]{Lewis98}, which,
  however, does not give an indication, why the contribution over the
  halfcircles vanish. In our case, these follow from the well-known
  asymptotics (see~\cite{Olver, SidiHoggan11}) for the modified Bessel function with respect to order,
  \[
    I_\lambda(z/2) \asymp \frac{(z/4)^\lambda}{\Gamma(\lambda + 1)}.
  \]

  The small time asymptotic behavior follows directly from the fact
  that the Laplace transform $g(\cdot,z)$ is a rapidly decreasing
  function. The large time asymptotic behavior follows from Watson's
  Lemma, since
  \[
  \int_0^\infty \frac{8s\sinh(\pi s)}{4s^2 +
  1} K_{is}(\gamma) \exp(-\mbox{$\frac12$}s^2 T') \dd s
  = 
  \int_0^\infty f(u) \exp(-u T') \dd u
\]
where 
\[
  f(u) = \frac{8\sinh(\pi\sqrt{2u})g(\sqrt{2u})}{1 + 8u}
  = 8\pi k(0) \sqrt{2u} ( 1 + k'(0)/k(0) \sqrt{2u} + \mathcal O (u))
\]
and $k(u) = K_{iu}(\gamma)$.

\end{proof}

\end{document}